\documentclass[11pt,a4paper]{article}
\pdfoutput=1
\usepackage{jheppub}
\usepackage{amsmath,amstext,amssymb}
\usepackage{graphicx}
\usepackage{feynmp}
\usepackage{caption}
\usepackage{float}
\begin{document}
\title{\bf Pomeron fan diagrams in perturbative QCD}
\author{J.Bartels$^a$, M.A.Braun$^b$\\
{\it $^a$ II.Institut fuer Theoretische Physik, Universitaet Hamburg, Luruper Chaussee 149,
D-22761 Hamburg.Germany}\\
{\it $^b$Dep.of High Energy Physics, Saint-Petersburg State University, \\198504 S.Petersburg,Russia}}

\begin{abstract}
{Within QCD reggeon field theory we study the formation of two
subsequent triple pomeron vertices in the process
P$\to$PP$\to$PPP. We make use of an earlier investigation
~\cite{Bartels:1999aw} of the six-reggeon amplitude in deep
inelastic scattering and show that in the large-$N_c$ limit
pomeron fan diagrams emerge with the same triple pomeron vertex in all places. We thus confirm the 
BK-equation, but we also find additional terms related to the reggeization of the gluon, and we discuss their potential significance. Our analysis also includes the general pomeron $\to$ two odderon vertex:
a particular version of this vertex has been included into earlier generalizations the BK equation.}
\end{abstract}

\maketitle

\input epsf

\def\beq{\begin{equation}}
\def\eeq{\end{equation}}
\def\bea{\begin{eqnarray}}
\def\eea{\end{eqnarray}}

\def\disc{{\rm Disc}}
\def\lra{\leftrightarrow}
\def\ep{\epsilon}
\def\hotimes{\hat{\otimes}}
\def\ot{\leftarrow}

\section{Introduction}

\begin{figure}
\hspace*{30 pt}
\epsfig{file=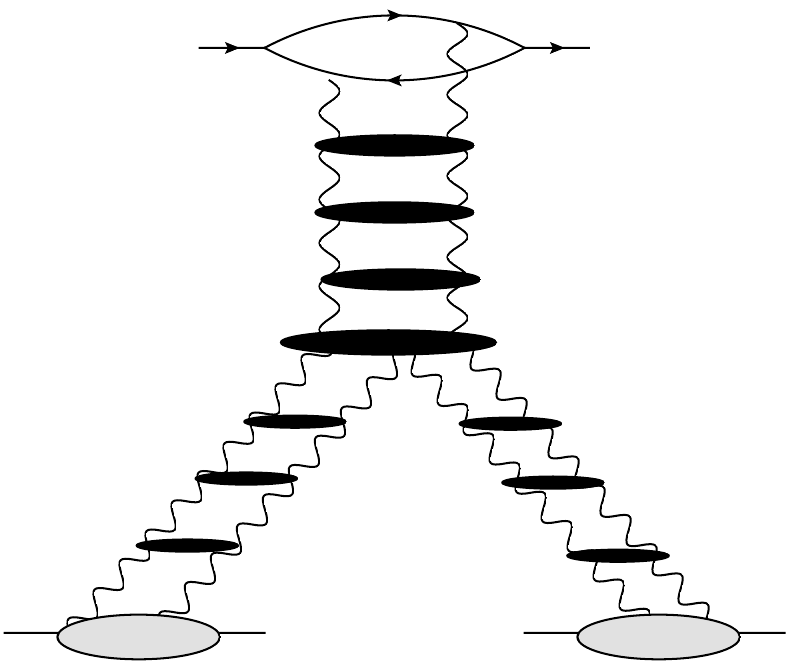, width=6cm\hspace*{50 pt}}
\caption{The triple pomeron vertex}
\label{fig1}
\end{figure}

Rather long time ago in the study of the interaction of a
colorless
projectile with two colorless targets the triple-pomeron vertex
$\Gamma$
was constructed, both within QCD reggeon field theory
\footnote{In our definition,
QCD reggeon field theory is defined as the theory of interacting
reggeized gluons. The BFKL pomeron appears as a bound state of
(at least) two reggeized gluons.}
~\cite{Bartels:1993ih,Bartels:1994jj} and
within the dipole picture ~\cite{mueller}. In Fig. \ref{fig1} we
illustrate the basic process in which the triple pomeron vertex
appears: the scattering of a projectile (virtual photon) on two
targets, described by the interaction of three BFKL ladders
which are coupled to the triple pomeron vertex.

\begin{figure}
\hspace*{140 pt}
\epsfig{file=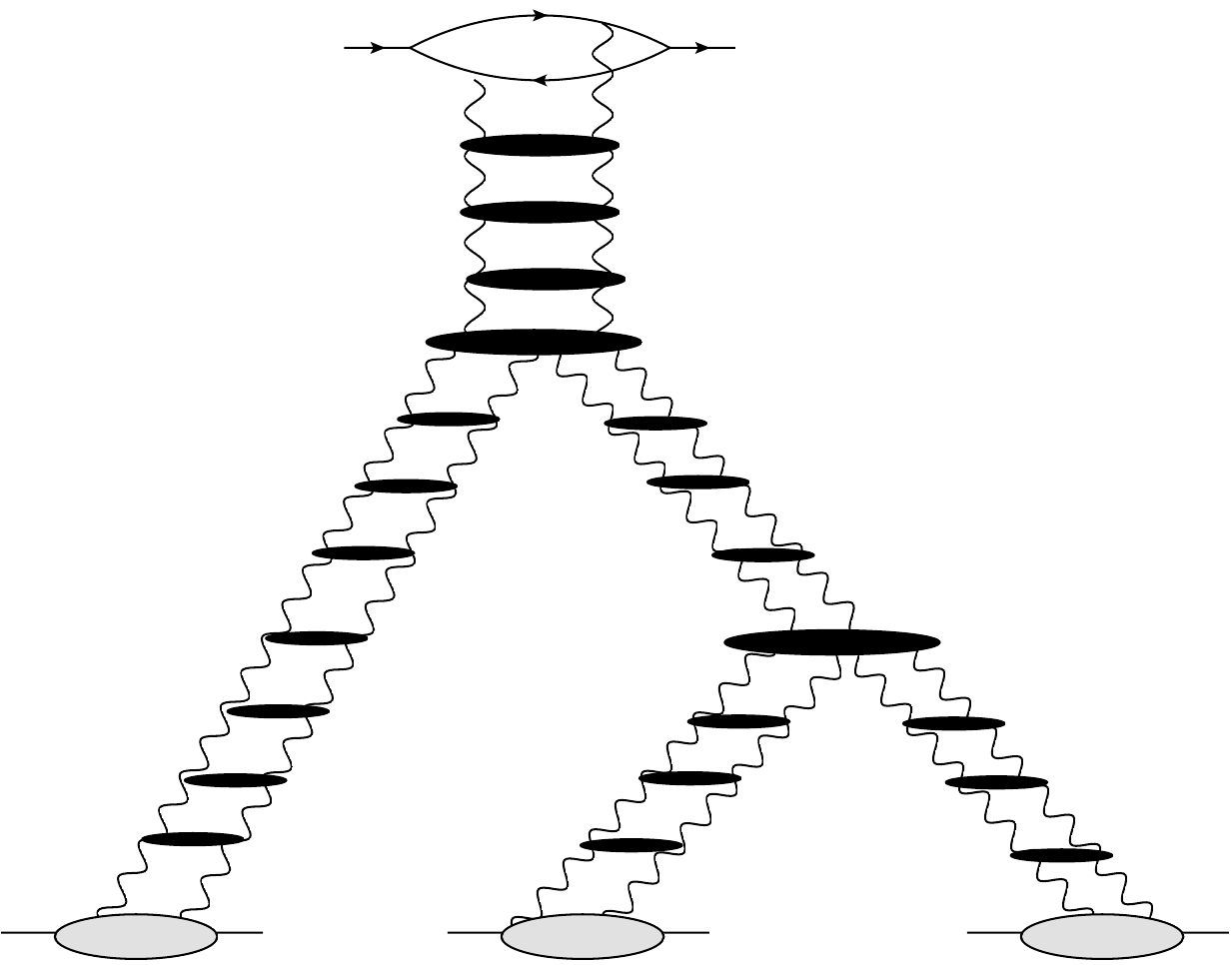, width=6 cm}
\caption{pomeron fan diagrams}
\label{fig2}
\end{figure}

\begin{figure}
\hspace* {160 pt}
\epsfig{file=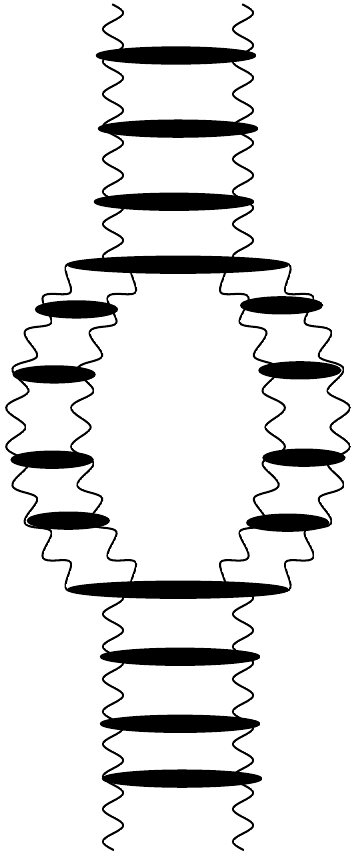, width=3 cm}
\caption{pomeron self-mass}
\label{fig3}
\end{figure}

In momentum space the triple pomeron vertex is given by a somewhat
lengthy function $V_{4 \ot 2}^{a_1a_2a_3a_4|b_1b_2}(k_1,k_2,k_3,k_4|q_1,q_2)$  \cite{Bartels:1999aw} which depends upon the color and  the transverse momenta of the constituent reggeons of the incomjng pomeron and the two outgoing pomerons. Switching via Fourier transform to the space of transverse coordinates and taking the large-$N_c$ limit, the expression takes the much simpler simple form
\cite{Braun:1997nu,Bartels:2004ef}:
\bea
&&\int d^2r_1d^2r_2d^2r_3 P(r_1,r_2) P(r_2r_3)  \Gamma(r_1,r_2,r_3)P(r_1,r_3)\nonumber\\
&&=-\frac{g^4N_c}{4\pi^3}\int d^2r_1d^2r_2d^2r_3
P(r_1,r_2) P(r_2r_3) 
\frac{r_{13}^2\nabla_1^2\nabla_3^2}
{r_{12}^2r_{23}^2}P(r_1,r_3).
\label{vdef}
\eea
Here  $P(r_1,r_3)$ denotes the (upper) incoming pomeron,  $P(r_1,r_2)$ and $P(r_2,r_3)$ the two (lower) outgoing pomerons.
They depend upon the transverse coordinates of their constituent reggeons, and  $r_{ik}=r_i-r_k$
(we have suppressed the color labels and the rapidity dependence of the pomerons).
This triple pomeron vertex plays an important role in
applications, since it serves as a fundamental building block in
constructing a theory of interacting pomerons,
which is expected to describe strong interactions in the Regge
kinematics.
In particular, it lies at the origin of the structure functions
of the heavy nuclei, as given by the Balitski-Kovchegov equation
~\cite{Balitsky:1995ub,Kovchegov:1999yj} which sums pomeron fan
diagrams  (Fig. \ref{fig2}). In this
nonlinear evolution equation  this vertex
represents the nonlinear part of the kernel.
It is also responsible for pomeron loops, in particular the
pomeron the self-mass (Fig. \ref{fig3}), which was calculated in
~\cite{bar2,pech, bra, bratar}.
In this calculation also the inverted vertex PP$\to$P was
introduced.

In the framework of QCD reggeon field theory, reggeon
interaction vertices are computed as a sum of several
contributions. As an example, the triple pomeron vertex is
obtained from the transition vertex: $2 \to 4$ reggeized gluons.
This interaction vertex is the sum of the integral kernel $K_{2
\to 4}$ plus also other 'induced pieces' which, at the end,
leads to the conformal invariant expression $V_{4 \ot 2}$. The
derivation of this vertex was an outcome
of the calculation of the high energy behavior of the six point
function (more precisely: from energy discontinuities of the
corresponding scattering amplitude). This six point amplitude
contains the triple pomeron vertex, as the single splitting of a
pomeron in two, illustrated in Fig. \ref{fig1}. The obtained
result is valid in all orders $N_c$, and the equivalence, in the
large-$N_c$ limit, with (\ref{vdef}) was proven in
\cite{Bartels:2004ef}.

In order to catch up with the fan diagrams of the BK-equation,
which have been derived in the color dipole approach, one needs
to consider more than one consecutive splitting: the simplest example is shown
Fig. \ref{fig2}. Within QCD reggeon field theory, this requires
the study of amplitudes containing al least 6 reggeized gluons at the lower end,
i.e.
the scattering of a projectile on three color singlet targets.
In earlier demonstrations
~\cite{Braun:2000wr} it was tacitly assumed that in consecutive
splittings exactly the same vertices $\Gamma$ appear. An explicit
derivation, however, is still missing. Also,
it will be interesting to what extent the fan structure is
complete and will be valid also beyond the large-$N_c$ limit. It
is the purpose of this study to fill some of these gaps.

An investigation of amplitudes with up to six reggeized gluons at the lower end
has been presented in \cite{Bartels:1999aw}. The results, in
principle, allow to find, in the leading-log approximation, the
complete structure of the evolution of the 2,4, and 6 gluon
systems.
However, the final step of analyzing the transition $4 \to 6$
gluons has not been completed. In the present paper we apply the
large-$N_c$ limit and show that, in fact,
the fan structure with two consecutive triple pomeron vertices
appears, in full agreement with the BK equations. There are, however, two features 
which do not affect the validity of the fan structure but go beyond the well-known results. First, we find 
additional contributions related to the reggization of the gluon. We will dicuss their potential significance. Next, our derivation includes the vertex: pomeron $\to$ two odderons \cite{Bartels:1999aw}. As we will discuss in more detail, this vertex is slightly more general than the one used in extensions of the BK equation \cite{kovchnew,itakura}. 

Our paper is organized as follows. We first (section 2)
recapitulate the main steps of the derivation of the triple
pomeron vertex in the context of the scattering
amplitude with three or four gluons at the lower end. We will restrict ourselves to the most essential
steps and, for details, refer to the literature. We then (section 3)
summarize the main results for the amplitude with five or six  gluons at the lower end
obtained in \cite{Bartels:1999aw}, and in particular we comment on the pomeron $\to$ two odderon vertex. In section 4 we concentrate on the large-$N_c$ limit and 
derive the fan structure where the same triple pomeron
vertex describes the consecutive splitting of two BFKL pomerons. In addition, we find the reggeizing pieces and in a separate subsection discuss their potential significance.
In a summarizing section we discuss our results in comparison with the BK equation.

\section {The $2 \to 4$ reggeized gluon transition: the
triple-pomeron vertex $V_4$}

To formulate the problem and to fix our notations we first
recall the main steps of the derivation of the triple -pomeron
vertex $V_{4 \ot 2}$ in QCD reggeon field theory,
the theory of interacting reggeized gluons. It will be useful to
recapitulate both the
original derivation \cite{Bartels:1993ih} (which is valid in all
orders $N_c$) and the large-$N_c$ ($N_c>>1$) derivation in
\cite{Braun:1997gm} which makes use of the
cylinder topology of the pomeron. 

\subsection {Two and three gluon states  coupled to the $q\bar{q}$ loop}
\begin{figure}
\hspace* {120 pt}
\epsfig{file=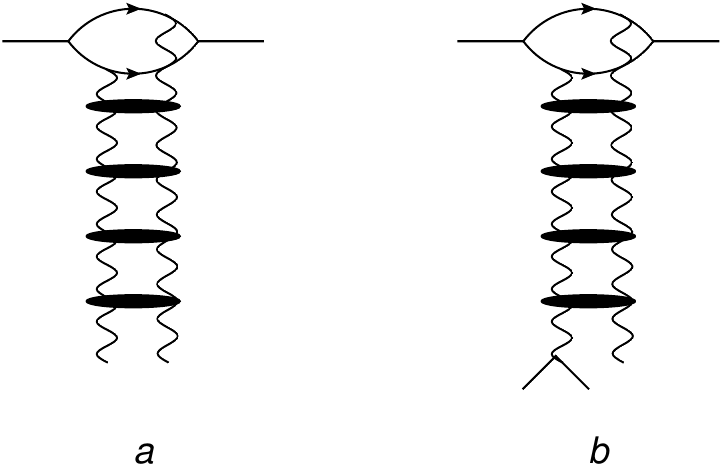, width=6 cm}
\caption{(a) the amplitudes $D_2$ and  (b) $D_3$}
\label{fig4}
\end{figure}

The structure of  intermediate states coupled to the upper fermion loop in Fig. \ref{fig1}
has long been known ~\cite{Braun:1997gm}.
We begin with the simplest case, $D_2$ in Fig.\ref{fig4}a, which describes a single BFKL pomeron coupled to the fermion loop.
The BFKL pomeron is a bound state of two reggeized gluons, and  $D_2$ satisfies the equation
\beq
S_2D_2=D_{20}+V_{12} D_2.
\label{d2}
\eeq
Here
$S_2$ is the two-reggeon free Schroedinger operator for energy $j_1=1-j$
\beq
\label{s2}
S_2=j_1+\omega(1)+\omega(2),
\eeq
where $j$ is the angular momentum, $\omega(i)$  the reggeon
Regge trajectory with momentum $q_i$, and $V_{12}$ the BFKL kernel written as an operator in the transverse momentum space (with the factor $g^2N_c$ included). Both $\omega(i)$ and
$V_{12}$ are well known and can be found in the literature.
$D_{20}(1,2)\equiv D_{20}(q_1,q_2)$ is the momentum dependent part of the impact factor
describing the four different couplings of the two reggeons 1 and 2 to the fermion loop.
With the color dependence included we have
\beq
D_{20}^{a_1a_2}(1,2)=\delta_{a_1a_2}D_{20}(1,2).
\eeq
$D_2$ denotes the full BFKL pomeron coupled to the impact factor (including angular momentum $j$ and both color and transverse momentum).

Here and in the following we adopt, for simplicity, the following notation: 
the product of two operators includes the integration over transverse  momenta, e.g.
\beq
V_{12}D_2: =\frac{g^2N_c}{8\pi^2} \int d^2 k V_{12}(k_1,k_2|k_1+k, k_2-k) D_2(k_1+k,k_2-k).
\eeq
Note that we define the operator $V_{12}$ in such a way that includes the rhs momentum propagators
$\frac{1}{(k_1+k)^2 (k_2-k)^2}$. 

Formally we can solve (\ref{d2}) by introducing the BFKL Green's-function $G_2$ which describes the infinite sum of ladder graphs. Sometimes it is convenient to use, instead of angular momentum $j$, rapidity $y$. We put
\beq
S_2= \partial_y+\omega(1)+\omega(2)
\eeq
and obtain the rapidity dependent Green's function $G_2(y)$.  We then can write the solution as a function of $y$:
\beq
D_2(y)=G_2(y) D_{20},
\eeq
where $G_2(y)$ satisfies the equation
\beq
(S_2- V_{12}) G_2(y) = \delta(y).
\eeq

Now consider the coupling of three reggeized gluons to the fermion loop. It is important to note that the three gluons at the lower end may either come from a direct coupling the fermion loop or from transitions from two reggeons coupled to the loop, which afterwards transform
into three reggeons by means of the splitting kernel $K_{3 \ot 2}$  \footnote{From now on we will find it convenient to orient, in the subscripts of the operators, the arrows from right to left. This is consistent with reading our equations which contain products of operators (beginning with (\ref{d2})) from right to left. In our figures, this corresponds to moving down from top to bottom.}. As a result, the equation for the corresponding amplitude $D_3$ contains two inhomogeneous terms:
\beq
S_3D_3=D_{30}+D_{3\leftarrow 2}+\frac{1}{2}(V_{12}+V_{23}+V_{31})D_3.
\label{d3}
\eeq
Here $D_{30}$ denotes the fermion loop with three reggeons coupled to it in all possible ways. Explicitly
\beq
D_{30}^{a_1a_2a_3}(1,2,3)=-\frac{1}{2}f^{a_1a_2a_3}g\Big(D_{20}(2,31)-D_{20}(1,23)-D_{20}(3,12)\Big).
\label{d30}
\eeq
Here '12' is our short-hand notation for the sum of the two momenta $q_1+q_2$.
The transition from 2 to 3 reggeons,  $D_{3\leftarrow 2}$, is accomplished by the kernel $K_{3\ot 2}$
\beq
D_{3\leftarrow 2}(y)=K_{3\leftarrow 2} D_2(y)=K_{3\leftarrow 2}G_2(y) D_{20},
\label{d2to3}
\eeq
where the momentum dependent part of the $2 \to 3$ kernel (including the momentum propagators  for the
intermediate two reggeon state) has the form:
\bea
&&K_{3\ot 2}(k_1,k_2,k_3|q_1,q_2)=\nonumber\\
&&=g^3N_c\Big\{
\frac{q_{12}^2}{q_1^2q_2^2}+\frac{k_2^2}{(q_1-k_1)^2(q_2-k_3)^2}-
\frac{k_{12}^2}{q_1^2(q_2-k_3)^2}-\frac{k_{23}^2}{q_2^2(q_1-k_1)^2}\Big\}.
\eea
It is remarkable that the solution of the equation (\ref{d3}) is quite simple, and can totally expressed via the pomeron wave function $D_2$ \cite{Bartels:1993ih} (Fig.\ref{fig4}b):
\beq
D_3^{a_1a_2a_3}(1,2,3)=-\frac{1}{2}gf^{a_1a_2a_3}g\Big(D_2(2,31)-D_2(1,23)-D_2(12,3)\Big).
\label{d31}
\eeq
This solution has the same momentum structure as the one-loop impact factor (\ref{d30}). Both (\ref{d31}) and (\ref{d30}) have the cylinder topology. 

The three terms on the rhs of (\ref{d31}) have a simple interpretation. As an example, the third term,
$D_2(12,3)$ illustrated in Fig.\ref{fig4}b), describes a single BFKL pomeron, where the left constituent reggeon carries the sum of two momenta, $q_1+q_2$, whereas the constituent reggeon on the rhs carries the momentum $q_3$. If the $q_i$ refer to targets 1,2,3 to which the pomeron couples, the left reggeon splits into two gluons which then couple to target $1$ and target $2$. Here it is important to note the difference between wavy lines (which correspond to reggeized gluons) and straight lines (which denote elementary gluons): the former ones 
contain evolution in rapidity whereas the latter ones do not. Fig.\ref{fig4}b) thus illustrates the nature of reggeization: a reggeized gluon can be viewed as a bound state of two elementary gluons.  Later on we will see that a reggeized can also be viewed as a bound state of three or even more gluons.          

\subsection{The triple pomeron vertex}

We now extend the procedure to transitions into four reggeized gluons. Let us first follow the derivation described in \cite{Bartels:1993ih}. The coupling of four reggeized gluons to the fermion loop consists of two groups:
\bea
\label{d40}
&&D_{40}^{a_1a_2a_3a_4}(1,2,3,4)\\
&&=-g^2d^{a_1a_2a_3a_4} \left(D_{20}(123,4) +D_{20}(1,234) -D_{20}(14,23)\right) \nonumber\\
&&-g^2d^{a_2a_1a_3a_4} \left(D_{20}(134,2) +D_{20}(124,3) -D_{20}(12,34)
-D_{20}(13,24)
\right), \nonumber
\eea
where the color tensor is defined as
\bea
d^{a_1a_2a_3a_4}&=&{\rm Tr}\,(t^{a_1}t^{a_2}t^{a_3}t^{a_4}+t^{a_4}t^{a_3}t^{a_2}t^{a_1}).
\nonumber\\
&=& \frac{1}{2N_c} \delta_{a_1a_2} \delta_{a_3a_4} + \frac{1}{4} \left(d_{a_1a_2c} d_{ca_3a_4} +f_{a_1a_2c} f_{ca_3a_4} \right).
\label{even-trace}
\eea
\begin{figure}
\hspace*{50pt}
\epsfig{file=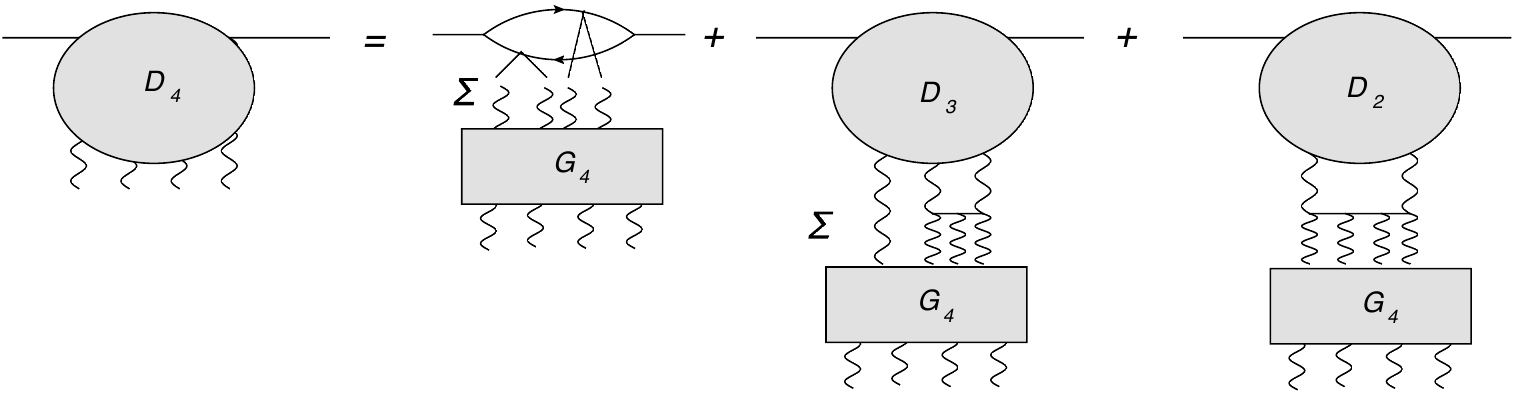, width=12cm,height=4cm}
\caption{Amplitude with 4 reggeized gluons. $G_4$ denotes the Green's function for 4 reggeized gluons.}
\label{fig5}
\end{figure}
By drawing all terms on the surface of a cylinder, the two groups differ in the order of $t$-channel gluons.
The integral equation for $D_4$ can be written as
\beq
S_4 D_4 = D_{4;0} +\sum K_{3 \ot 2}D_{3} +K_{4 \ot 2} D_2 +\sum_{(ij)} V_{ij} D_4,
\label{eqd4}
\eeq
where the momentum dependent part of the kernel $K_{4\ot 2}$ has the form:
\bea
&&K_{4 \ot 2}(k_1,k_2,k_3,k_4|q_1,q_2)\nonumber\\
 &=& gK_{3\ot 2}(k_1,k_{23},k_4|q_1,q_2) \nonumber\\
&=&g^4N_c \Big\{
\frac{q_{12}^2}{q_1^2q_2^2}+\frac{(k_2+k_3)^2}{(q_1-k_1)^2(q_2-k_4)^2}-
\frac{k_{123}^2}{q_1^2(q_2-k_4)^2}-\frac{k_{234}^2}{q_2^2(q_1-k_1)^2}\Big\}.
\eea
Formally we could solve  (\ref{eqd4}) by
\beq
D_4(y)=G_4(y)  D_{40} +  \int dy' G_4(y-y') \Big[\sum K_{3 \ot 2}D_{3}(y') +K_{4 \ot 2} D_2(y')\Big].
\label{sol-eqd4}
\eeq
This solution is illustrated in Fig.\ref{fig5}.

\begin{figure}
\hspace*{50pt}
\epsfig{file=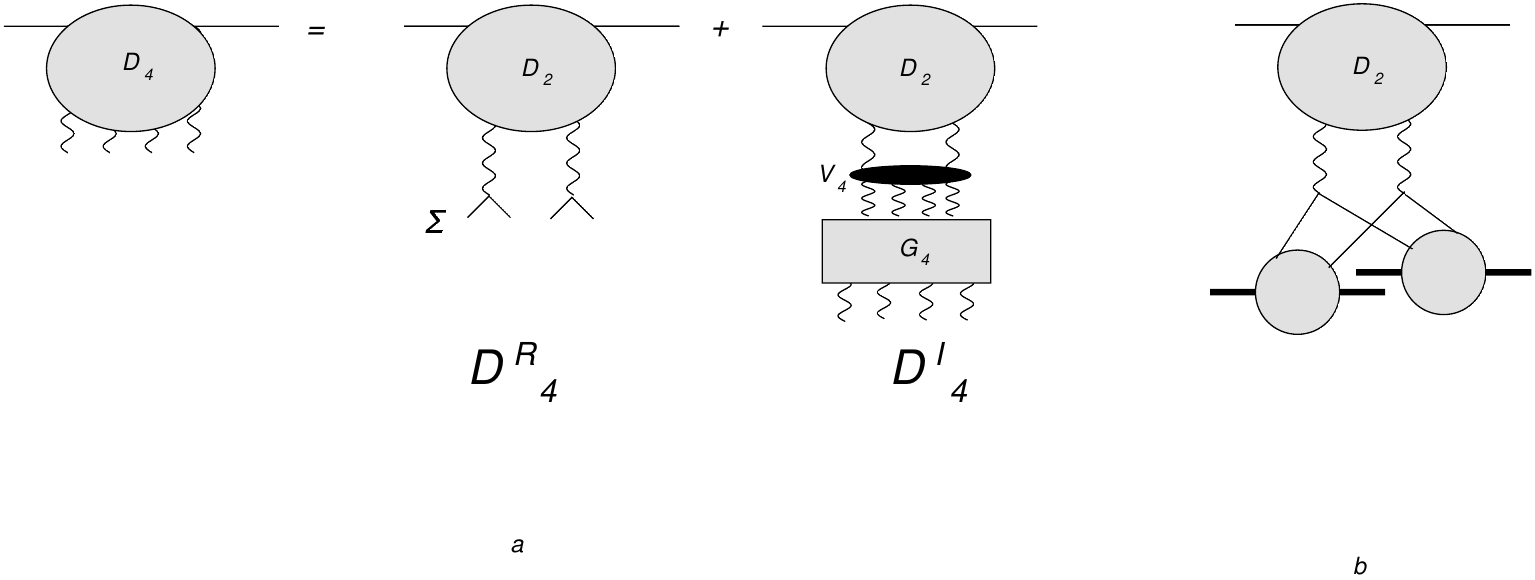, width=12cm,height=4cm}
\caption{(a) Amplitude with 4 reggeized gluons: decomposition into reggeized (left) and irreducible (right)
pieces. The detailed form of the reggeized part, $D_2$, is given in \ref{d4R}). It is the 
just the BFKL pomeron where each lower leg carries sums of momenta and finally splits into elementary  gluons. (b) Example of the coupling of a reggeized term to two nucleons.}
\label{fig6}
\end{figure}

However, this is not the final solution. Instead, one first removes a reggeized piece with only two reggeized gluons, i.e. one  makes the ansatz (Fig.\ref{fig6})
\beq
D_4=D_4^R+D_4^I
\label{reduction1}
\eeq
 with
\bea
\label{d4R}
&&D_{4}^{R ;a_1a_2a_3a_r}(1,2,3,4)=\nonumber\\
&&-g^2d^{a_1a_2a_3a_4} \left(D_{2}(123,4) +D_{2}(1,234) -D_{2}(14,23)\right)\nonumber\\
&&-g^2d^{a_2a_1a_3a_4} \left(D_{2}(134,2) +D_{2}(124,3) -D_{2}(12,34)
-D_{2}(13,24)
\right). 
\eea
Inserting this into (\ref{eqd4}) and using the integral equation for $D_2$, eq.(\ref{d2}), we
combine all terms containing $D_2$ and arrive at the equation for $D_4^I$
\beq
S_4 D_4^I = V_{4 \ot2}  D_2 +  \sum_{(ij)} V_{ij} D_4^I
\label{eqd4I}
\eeq
with the solution
\beq
\label{sol-eqd4I}
D_4^I(y) = \int dy' G_4(y-y')  V_{4 \ot 2} D_2(y').
\eeq

Here $V_{4 \ot2} $ denotes the triple pomeron vertex as an operator in the color and transverse momentum space.
Exhibiting the color structure, this
vertex can be written as a sum of three terms:
\bea
&&V_{4 \ot 2}^{a_1a_2a_3a_4}(k_1k_2k_3k_4|q_1,q_2) )= \nonumber\\
&&=\delta_{a_1a_2}\delta_{a_3a_4}V_4(1,2;3,4)+
\delta_{a_1a_3}\delta_{a_2a_4}V_4(1,3;2,4)+
\delta_{a_1a_4}\delta_{a_2a_3}V_4(1,4;2,3),
\label{d4tot}
\eea
where on the rhs we have suppressed the initial momenta $q_1,q_2$ which in the convolution with 
$D_2$ are to be integrated, and we have written only the momentum structure of the two outgoing pomerons with the substitutions 
$k_i \to i, (i=1,...,4)$.The function $V_4(1,2;3,4)$ is symmetric under
permuting $(1,2)$ or $(3,4)$, and under the exchange of the pairs $(1,2)$ and $(34)$; also, the convolution $V_4(1,2;3,4)D_2$
vanishes whenever one of the four transverse momenta goes to zero.  For the full vertex one observes Bose symmetry under the exchange of reggeized gluons $1,2,3,4$.
Considering the configuration where the pairs $(1,2)$ and $(34)$ are color singlet and taking the large-$N_c$ limit we find that in (\ref{d4tot}) only the first term contributes, and integral equation for $D_4^{I}$ simpifies:
\beq
S_4 D_4^{I}((1,2),(3,4))=(V_4 D_2) (12,34) +\left( V_{12}+V_{34}\right) D_4^{I},
\label{d4-largeNc}
\eeq
where we have indicated that, in both color and momentum, the arguments of $V_4 D_2$ and $D_{2}$ are grouped into the pairs $(1,2)$ and $(3,4)$.
Since the rapidity evolution of these pairs proceeds independently, the Green's function $G_4$ factorizes:
\beq
\label{G4-large-Nc}
G_4(y;1234) \to G_2(y;12) G_2(y;34),
\eeq
and $D_4^{I}$ depends upon two independent rapidity variables, $y_1$ and $y_2$:
\beq
\label{D4I-sim}
D_4^{I}(y_1,y_2;12,34)= \int dy'  G_2(y_1-y';12) G_2(y_2-y';34)( V_4 D_2(y')).
\eeq
After Fourier transformation to coordinate space, this expression takes the form (\ref{vdef}) (with suitable substitutions for the three pomerons). A proof (for general $N_ c$) of this Fourier tansform, has been presented in \cite{Bartels:2004ef}.

Finally we make a few comments on the reggeizing term $D_4^R$. As indicated in Fig.\ref{fig6}a. and described in (\ref{d4R}), the reggeized term comes as a sum of single BFKL pomerons where the lower legs of reggeized gluons split into four elementary gluons. $D_4^R$ represents an additive correction to the single pomeron in Fig.\ref{fig4}a with two gluons 
at the lower end. In the scattering on a big nucleus the pomeron in Fig.\ref{fig4}a, in its simplest form, couples to a single nucleon inside the nucleus, and this coupling can be described by the gluon distribution of the nucleon. Following this simple picture, the corrections contained in $D_4^R$ then can be interpreted as the coupling of a single BFKL pomeron to a pair  of two nucleons. As an example, consider the last term in 
(\ref{d4R}), $D_2(13,24)$ (remember that we are considering the configuration where the 
pairs $(12)$ and $(34)$ are in color singlet states). As illustrated in Fig.\ref{fig6}b, in this contribution the BFKL pomeron couples to two different nucleons. Since all the rapidity evolution is contained in the reggeized gluons, this coupling to 
two nucleons is a 'low energy' interaction. As we will discuss in section 4.4, this new contribution can be
interpreted as the second order term in an eikonal
initial condition for the nonlinear evolution equations \cite{braun:tmf}.

\subsection{The topological derivation}

As observed in \cite{Braun:1997gm}, the large-$N_c$ limit of the
triple pomeron vertex can be obtained also in another way, by
making use of the cylinder topology of the pomeron in the
large-$N_c$ approximation.

To this end we return to $D_{40}$ in (\ref{d40}) and note that
the two groups correspond to two different configurations on the
surface of the cylinder, $D_4^{1234}$ and $D_4^{2134}$:
\beq
D_{40}^{a_1a_2a_3a_4}(1,2,3,4)=-d^{a_1a_2a_3a_4}D_{40}^{1234}(1,2,3,4)-
d^{a_2a_1a_3a_4}D_{40}^{2134}(1,2,3,4),
\label{d40top}
\eeq
where
\beq
D_{40}^{1234}=g^2\Big(D_{20}(1,234)+D_{20}(123,4)-D_{20}(14,23)\Big),
\label{d401}
\eeq
\beq
D_{40}^{2134}=g^2\Big(D_{20}(2,134)+D_{20}(124,3)-D_{20}(12,34)
-D_{20}(13,24)\Big).
\label{d402}
\eeq
They differ in the order of gluons on the surface of a cylinder
(indicated by the upper index),
and they do not mix.
The evolution in rapidity of the four-reggeon-state can start directly from
the
fermion loop, as described in (\ref{d401}) and (\ref{d402}), or
from two- and three-reggeon states with subsequent transitions
to a four gluon state.
One finds that transitions from the two-reggeon states into
$D_4^{2134}$ are prohibited. So the equations are
\beq
S_4D_4^{1234}=D_{40}^{1234}+D_{4\ot 2}^{1234}+D_{4\ot 3}^{1234}+\frac{1}{2}(V_{12}+V_{23}+V_{34}+V_{41})D_4^{1234}
\label{d41}
\eeq
and
\beq
S_4D_4^{2134}=D_{40}^{2134}+D_{4\ot 3}^{2134}+
\frac{1}{2}(V_{21}+V_{13}+V_{34}+V_{42})D_4^{2134}.
\label{d42}
\eeq
Here the transition from 2 to 4 reggeons is described by
\beq
D_{4 \ot 2}^{1234}=- K_{4 \ot 2} D_2.
\eeq
For the transition $3 \to 4$ we have:
\[
D_{4\ot 3}=\frac{1}{2}\Big( K_{3\ot 2}(1,2,3|1'3')
D_3(1',3',4)
-K_{3\ot 2}(1,2,4|1',4')D_3(1',3,4')\]\beq+K_{3\ot
2}(2,3,4|2',4') D_3(1,2',4')-
K(1,3,4|1',4') D_3(1',2,4')\Big),
\label{inh3}
\eeq
where $D_3$ is the solution of Eq. (\ref{d3}).
As shown in \cite{Braun:1997gm} , Eqs. (\ref{d41}) and
(\ref{d42}) for $D_4$ can be solved directly: the solutions can
be obtained from (\ref{d401}) and (\ref{d402}) by
substituting
$D_{20} \to D_2$. Their sum coincides with $D_{4}^R$ in
(\ref{d4R}).

In the next step we consider another toplogical configuration,
which is suppressed in
$1/N_c$: the formation of two color singlet cylinders made of
reggeons pairs (1,2) and (3,4):
we denote this amplitude by $D_4^{2P}$.
These two cylinders can start directly from the fermion loop or
from a single cylinder formed by 2, 3 or 4 reggeon which then
couples to two cylinders.
The full amplitude $D_4^{2P}$
satisfies the equation ~\cite{Braun:1997gm}:
\beq
S_4D_4^{2P}=D_{2P0}+D_{2P\ot 2}+D_{2P\ot 3}+D_{2P\ot 4}+
(V_{12}+V_{34})D_4^{2P}.
\label{d42p}
\eeq
The inhomogeneous term $D_{2P0}$ is just the projection of
the loop $D_{40}$ onto the 2-pomeron state:
\beq
 D_{2P0}=\frac{1}{2}N_c(D_{40}^{1234}+D_{40}^{2134}).
\label{d2p0}
 \eeq
The inhomogeneous terms $D_{4\ot 2}^{2P}$, $D_{4\ot 3}^{2P}$ and
$D_{4\ot 4}^{2P}$
describe the
mentioned transitions into the two-pomeron state
from two, three and four evolved reggeons coming from the loop.
\beq
D_{2P\ot 2}=-K_{4\ot 2}D_2,
\label{inh2}
\eeq
where
\beq
K_{4\ot 2}(1,2,3,4|1',2')=gK_{3\ot 2}(1,23,4|1',2'),
\eeq
and $D_2$ is the solution of (\ref{d2}).
\[
D_{2P\ot 3}=\frac{1}{2}\Big( K_{3\ot 2}(1,2,3|1'3')
D_3(1',3',4)
-K_{3\to 2}(1,2,4|1',4') D_3(1',3,4')\]\beq+K_{3\ot
2}(2,3,4|2',4') D_3(1.2',4')-
K(1,3,4|1',4')D_3(1',2,4')\Big),
\label{inh3a}
\eeq
where $D_3$ is the solution of Eq.(\ref{d3}).
The transition from the single cylinder to the two-cylinder
configuration is
described by
\beq
D_{2P\ot 4}=(V_{23}+V_{14}-V_{23}-V_{24})
(D_4^{1234}-D_4^{2134}),
\label{inh4}
\eeq
where $D_4^{1234}$ and $D_4^{2134}$ are the solutions to Eqs.
(\ref{d41}) and (\ref{d42}).
In order to solve (\ref{d42p}) we collect all terms containing
$D_2$ and arrive at
(\ref{d4-largeNc}) with the same triple pomeron vertex function
$V_4$. Formally we can solve this equation:
\beq
D_{4}^{2P}(y)=\int_0^ydy_1dy_2\delta(y_1+y_2-y)G_{2P}(y_2)V_4
D_2(y_1),
\label{d4triple}
\eeq
where  $G_{2P}=G_2 \cdot G_2$ is the Green function for two non-interacting
pomerons.
For the complete four reggeon amplitude we still have to add
the single cylinder contribution $D_4^R$.

We finally want to stress that all results obtained so far do
not use the explicit form of function $D_{20}(1,2)$, and they
remain valid if one substitutes $D_{20}$ by an arbitrary
function $F_2(1,2)$ with properties
\beq
F_2(q_1,q_2)=F_2(q_2,q_1),\ \ F_2(0,q_2)=F_2(q_1,0)=0.
\label{eq2}
\eeq
All one needs is that the three and four-reggeon contributions
come from the impact
factors, $F_3$ and $F_4$, related to $F_2$ in exactly the same
manner as $D_{30}$
and $D_{40}$ are related to $D_{20}$ in Eqs. (\ref{d30}) and
(\ref{d40}), namely
\beq
F_3^{a_1a_2a_3}(1,2,3)==-\frac{1}{2}f^{a_1a_2a_3}g\Big(F_2(2,31)-F_2(1,23)-F_2(3,12)\Big)
\label{eq3}
\eeq
and
\[
F_4^{a_1a_2a_3a_4}(1,2,3,4)=-g^2\Big\{d^{a_1a_2a_3a_4}\Big(F_2(1,234)+F_2(123,4)-F(14,23)\Big)\]
\beq+
d^{a_2a_1a_3a_4}\Big(F_2(2,134)+F_2(124,3)-F_2(12,34)
-F_2(13,24)\Big)\Big\}.
\label{eq4}
\eeq

\section{The transition $2\to6$}

Let us now turn to the amplitudes with five and six reggeized
gluons at the lower end. For the time being
we postpone the
large-$N_c$ limit and keep the number of colors finite. In
\cite{Bartels:1999aw} an integral equation for $D_6$ has been
derived and analyzed. The final step, however, the analysis of
the irreducible piece $D_6^I$, has not been completed.
In section 4  we return to the large-$N_c$ limit, and we  will show that in this limit $D_6^I$,
contains the fan diagrams entering the Balitsky-Kovchegov (BK) equation.

\begin{figure}
\hspace*{0pt}
\epsfig{file=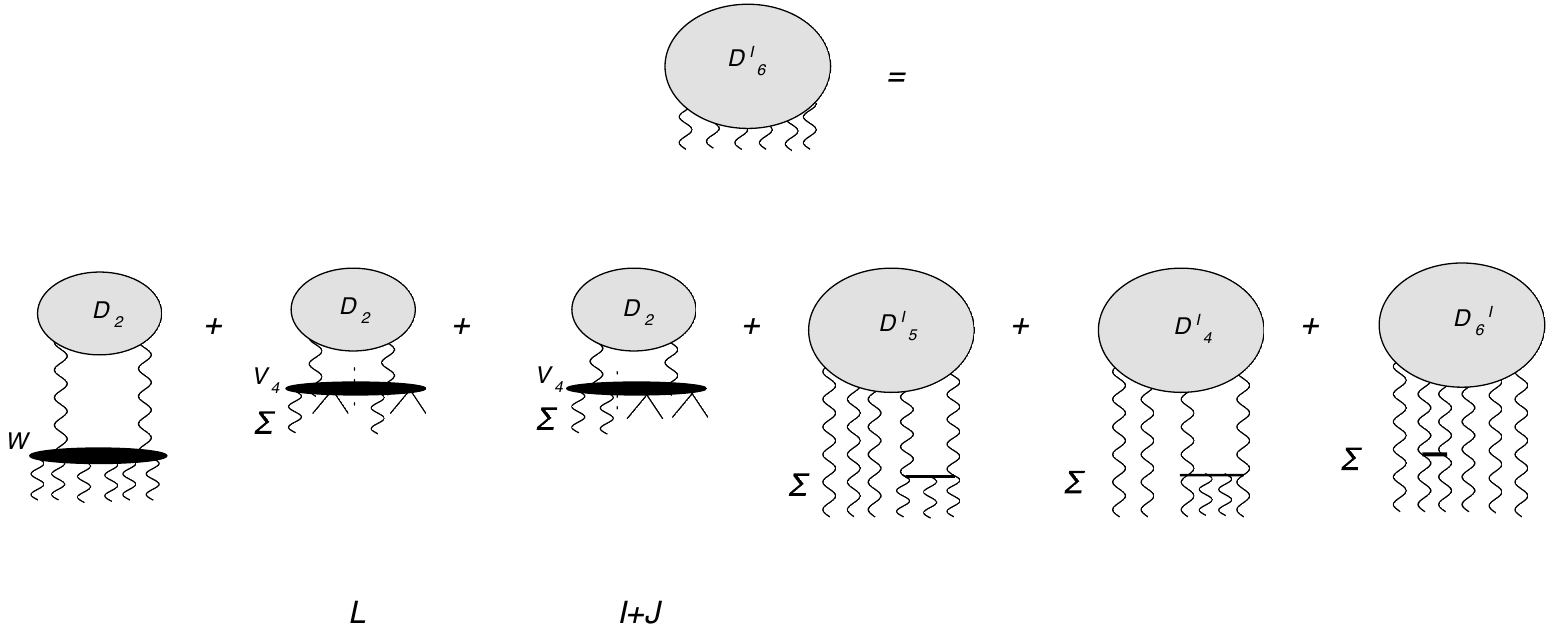,width=15cm,height=6cm}
\caption{The integral equation (\ref{d6I}) for the amplitude $D^{I}_6$ with
6 reggeized gluons.  In the terms denoted by 'L' and 'I+J'  the pairs of straight lines attached to the vertices denote reggeons which carry sums of final momenta.
For further explanations see text.}
\label{fig7}
\end{figure}

Let us first review the results obtained in
\cite{Bartels:1999aw}.  The
structure of the integral equation is illustrated in Fig. \ref{fig7}. The
analysis described in the previous section has illustrated that
the functions $D_i$ exhibit the following hierarchy structure:
$D_3$ is built from
$D_2$, $D_4$ is based upon $D_2$ and $D_3$. Similarly, $D_5$
contains $D_2$, $D_3$, and $D_4$, and $D_6$ is obtained from all
$D_i$ with $i<6$. This is illustrated by the integral equation
which, in a somewhat simplified form, reads as follows::
\beq
S_6 D_6=D_{60}+K_{6 \ot 2}D_2+ \sum K_{5 \ot 2}D_3 + \sum K_{4 \ot
2} D_4 + \sum K_{3 \ot 2}D_5 + \sum V_{ij} D_6
\label{d6}
\eeq
with
\bea
&&D_{60}^{a_1a_2a_3a_4a_5a_6}(1,2,3,4,5,6)=\nonumber\\
&&g^6\Big\{ d^{a_1a_2a_3a_4a_5a_6}
\Big[D_{20}(12345,6)+D_{20}(1,23456)-D_{20}(16,2345)\big]\nonumber\\
&&+d^{a_2a_1a_3a_4a_5a_6}
\Big[D_{20}(1345,62)-D_{20}(1345,62)+D_{20}(126,345)-D_{20}(12,3456)\big]\nonumber\\
&&+d^{a_1a_2a_3a_4a_6a_5}
\Big[D_{20}(12346,5)-D_{20}(1234,56)+D_{20}(156,234)-D_{20}(15,2346)\big]\nonumber\\
&&+d^{a_2a_1a_3a_4a_6a_5}
\Big[-D_{20}(1256,34)-D_{20}(1346,25)+D_{20}(125,346)+D_{20}(134,256)\big]\nonumber\\
&&+d^{a_3a_1a_2a_4a_5a_6}
\Big[D_{20}(12456,3)-D_{20}(1245,36)+D_{20}(136,245)-D_{20}(13,2456)\big]\nonumber\\
&&+d^{a_1a_2a_3a_5a_6a_4}
\Big[D_{20}(12356,4)-D_{20}(1235,46)+D_{20}(146,235)-D_{20}(14,2356)\big]\nonumber\\
&&+d^{a_2a_1a_3a_5a_6a_4}
\Big[-D_{20}(1246,35)-D_{20}(1356,24)+D_{20}(124,356)+D_{20}(135,246)\big]\nonumber\\
&&+d^{a_1a_2a_3a_6a_5a_4}
\Big[-D_{20}(1236,45)-D_{20}(1456,23)+D_{20}(123,456)+D_{20}(145,236)\big]\Big\},\nonumber\\
\label{d60}
\eea
where
\beq
d^{a_1a_2a_3a_4a_5a_6}={\rm
Tr}\,(t^{a_1}t^{a_2}t^{a_3}t^{a_4}t^{a_5}t^{a_6})+{\rm
Tr}\,(t^{a_6}t^{a_5}t^{a_4}t^{a_3}t^{a_2}t^{a_1}).
\eeq
One easily verifies that all configurations can be drawn on the
surface of a cylinder,
and they represent the different orderings of the $t$-channel
gluon lines.

On the rhs of (\ref{d6}), a new ingredient appears, $D_5$, for
which the following decomposition has been found:
\beq
D_5=D_5^R + D_5^I,
\eeq
where
\bea
\label{d5R}
D_5^{Ra_1a_2a_3a_4a_5}(1,2,3,4,5)=-g^3 \Big\{
f^{a_1a_2a_3a_4a_5} \Big[
D_2(1234,5)+D_2(1,2345)-D_2(15,234)\Big]\nonumber\\
+f^{a_2a_1a_3a_4a_5} \Big[
D_2(1345,2)-D_2(12,345)+D_2(125,34)-D_2(134,25)
\Big]\nonumber\\
+f^{a_1a_2a_3a_5a_4} \Big[
D_2(1235,4)-D_2(14,235)+D_2(145,23)-D_2(123,45)
\Big]\nonumber\\
+f^{a_1a_2a_4a_5a_3} \Big[
D_2(1245,3)-D_2(13,245)+D_2(135,24)-D_2(124,35)
\Big]\Big\}\nonumber\\
\eea
with
\beq
f^{abcde}=\frac{1}{i} \Big[{\rm Tr}\,(t^a t^b t^c t^d t^e)- {\rm
Tr}\,(t^e t^d t^c t^b t^a)\Big]
\eeq
and
\bea
D_5^{Ia_1a_2a_3a_4a_5}(1,2,3,4,5)= \frac{g}{2}
\Big[ f^{a_1a_2c}D_4^{Ica_3a_4a_5}(12,3,4,5)+
f^{a_1a_3c}D_4^{Ica_2a_4a_5}(13,2,4,5)+\nonumber\\
f^{a_1a_4c}D_4^{Ica_2a_3a_5}(14,2,3,5)+
f^{a_1a_5c}D_4^{Ica_2a_3a_4}(15,2,3,4)+\nonumber\\
f^{a_2a_3c}D_4^{Ia_1ca_4a_5}(1,23,4,5)+
f^{a_2a_4c}D_4^{Ia_1ca_3a_5}(1,24,3,5)+\nonumber\\
f^{a_2a_5c}D_4^{Ia_1ca_3a_4}(1,25,3,4)+
f^{a_3a_4c}D_4^{Ia_1a_2ca_5}(1,2,34,5)+\nonumber\\
f^{a_3a_5c}D_4^{Ia_1a_2ca_4}(1,2,35,4)+
f^{a_4a_5c}D_4^{Ia_1a_2a_3c}(1,2,3,45)\Big].\nonumber\\
\label{d5tot}
\eea
We remind that the triple pomeron vertex $V_{4 \ot 2}$ inside $D^{I
a_1a_2a_3a_4}_4$ has the structure described in (\ref{d4tot}).

Turning to the rhs of (\ref{d6}) and using our results for
$D_2$,..,$D_5$ we note that all terms (disregarding, for the
moment, the first term $D_{6;0}$), have a simple structure of
multi-reggeon states: the second term (containing $D_3$) and third term (containing $D_4$) start, at
the fermion loop, with a two-reggeon state which then turns into
the six-reggeon state. The fourth and fifth terms
have both 'R' and 'I' contributions; the 'R'-terms start with
two reggeon states and then undergo $2 \to 6$ transitions. The
'I'-terms begin with a two-reggeon state, then contain the $2\to
4$ transition described by the $2\to4$ reggeon vertex discussed
in the previous section, and finally end with a transition into
the six-reggeon state. This pattern is not altered if we insert,
for $D_6$, the decomposition 
\beq
D_6 = D_6^R + D_6^I
\label{decomp6}
\eeq
with $D_6^R$ being obtained from $D_{60}$ in (\ref{d60}) by
replacing on the rhs $D_{20} \to D_2$.
Using the BFKL equation for $D_6^R$ removes, on the rhs of
(\ref{d6}), the first term, $D_{60}$, and adds new
contributions to the transition $2 \to 6$ vertex. Additional
contributions to the $2 \to 6$ vertex are generated by
inserting, in the last term of the rhs of (\ref{d6}), $D_6^{R}$.
The main task
therefore is the computation of the complete (and potentially
new) $2\to6$ and $4\to6$ vertices.
In section 4.4. we will say more about the reggeizing term $D_6^R$.

A very big step was already done in \cite{Bartels:1999aw}.
Namely, starting from (\ref{d6}),
inserting (\ref{decomp6}) and performing an extensive
recombination of terms a modified equation for $D_6^I$ was
obtained which has the following structure:
\beq
S_6 D_6^I = W_{odd}D_2 + (T_L+T_I+T_J)D_2 + \sum K_{3 \ot 2}
D_5^I+
\sum K_{4 \ot 2} D_4^I + \sum V_{ij} D_6^I.
\label{d6I}
\eeq
This equation is illustrated in Fig.\ref{fig7}.

Here the first term denotes the $P \to 2 O$ pomeron-odderon
vertex, first derived and discussed
in ~\cite{Bartels:1999aw}. It has the form:
\beq
 W_{odd}D_2=\sum_{partitions} d^{a_1a_2a_3}d^{a_4a_5a_6}
\left(WD_2\right)(1,2,3;4,5,6),
\label{Odd-vertex}
\eeq
where the sum extends over all partitions of 6 reggeons into two
triplets, and $d^{abc}$
are the symmetric structure constants of the $SU(3)$ gauge
group. For details of the function $W$ we refer to
\cite{Bartels:1999aw}. 

Let us say few words about the significance of this vertex.
As it is well known, the odderon is a bound state of three reggeized gluons in a C-odd state, and it is natural that the pomeron$\rightarrow$two-odderon comes as a part of the 2 gluon $\rightarrow$ 6 gluon vertex. A special bound state solution of the three gluon system with intercept exactly at one was found in  \cite{Bartels:1999yt}. In this solution, two of the three gluons merge into a single even signature reggeized gluon (degenerate with the 'normal' odd-signature reggeized gluon) and then form a bound state with the third gluon. This results in a quasi-two-gluon bound state which is closely related to the BFKL pomeron. For this special  
odderon solution, the pomeron$\rightarrow$two-odderon vertex becomes similar to the triple pomeron vertex. It is this special odderon solution which appears in the generalizations of the BK-equation \cite{kovchnew,itakura}. Our more general version (\ref{Odd-vertex}), on the other hand,  allows to inlcude also 
other three-gluon solutions, such as the one found in \cite{wj,jw} with intercept slightly below one.

Returning to the equation (\ref{d6I}). the second term has the form:
\beq
T_L D_2=\sum_{partitions}
f^{a_1a_2a_3}f^{a_4a_5a_6}L(1,2,3;4,5,6),
\label{L-structure1}
\eeq
where the sum extends over all partitions of 6 reggeons in two
groups with three reggeons in each.
The function $L$ consists of $3^2=9$ terms which, for the partition
$(123)(456)$ can be written as follows:
\beq
L(1,2,3;4,5,6) =\frac{g^2}{4} \sum_{ijk} \sum_{lmn} (V_4D_2)
(ij,k;lm,n),
\label{L-structure2}
\eeq
where the sums are defined as follows
\bea
\sum_{ijk} = (12,3)+(1,23)-(13,2)\nonumber\\
\sum_{lmn}= (45,6)+(4,56)-(46,5).
\label{L-structure3}
\eea
Here each sum has a structure similar to $D_{30}$ in
(\ref{d30}). In the second term of Fig.\ref{fig7}, the
summation sign denotes both the summation over partitions
(\ref{L-structure1}) and the summations contained in
(\ref{L-structure2}). As it can be read off from
Fig.\ref{fig7}, these L-terms contain a transition described by $V_{4 \ot 2}$:
a pair of two reggeized gluons $\to$ splits into two colorless triplets of reggeized
gluons. For each triplet,  the initial 
momentum structure of the L-terms is similar to
$D_{30}$ in (\ref{d30}). Except for the large-$N_c$ limit, where each triplet evolves independently in rapidity, in the course of the subsequent evolution the triplets interactwith each other and thus loose their initial triplet structure.

Next we discuss the third term, which consists of two terms
labeled by 'I' and 'J'.
We first list the color structure:
\beq
T_I D_2=\sum_{partitions} \delta_{a_1a_2}
d^{a_3a_4a_5a_6}I(1,2;3,4,5,6)
\label{I-structure1}
\eeq
and
\beq
T_J D_2=\sum_{partitions} \delta_{a_1a_2} d^{a_4a_3a_5a_6}
J(1,2;3,4,5,6).
\label{J-structure1}
\eeq
In both terms we sum over all partitions of 6 reggeons into two
groups with 4 reggeons in one and 2 in the other one. In detail,
for the partition $(12)(3456)$:
\beq
I(1,2;3,4,5,6)=-g^2\Big(
(V_4D_2)(1,2;3,456)+(V_4D_2)(1,2;345,6)-(V_4D_2)(1,2;36,45)\Big)
\label{I-structure2}
\eeq
and
\bea
J(1,2;3,4,5,6)=-g^2\Big( (V_4D_2)(1,2;
356,4)+(V_4D_2)(1,2;346,5)\nonumber\\-(V_4D_2)(1,2;34,56)
-(V_4D_2)(1,2;35,46) \Big).
\label{J-structure2}
\eea
Focussing on the subsystem $(3456)$ and comparing the function
$I$ with the first line of $D_{40}$ in (\ref{d40}) one sees the
identical structure in the arguments; similarly, the four terms
in the function $J$ coincide with the second line in
(\ref{d40}). Following our discussion of the 'L'-term, we
interpret this structure as a transition:
a colorless pair of reggeized gluons $\to$ a colorless pair + a colorless quartet of reggeized
gluons. Initially, at the vertex $V_{4 \ot 2}$, the quartet has the same momentum structure
as $D_{40}$ in (\ref{d40}). During the subsequent evolution
the separation of pair+quartet will be lost.

Finally, the last two terms in Fig.\ref{fig7} still
contain 'elementary kernels', $K_{3 \ot 2}$ and $K_{4 \ot 2}$
which require further recombination. In the terms with $K_{3 \ot
2}$
and $K_{4 \ot 2}$ the sum extends over all pairs of gluons
inside $D_5^I$ and $D_4^I$, resp.

As we have said before, for general $N_c$ the missing next step has not been carried out. In the following we 
restrict ourselves to the large-$N_c$ limit and
describe the solution of eq.(\ref{d6I}) (and Fig.\ref{fig7}).

\section{The large-$N_c$ limit of $D_6$}
\subsection{New reggeizing pieces}

As already indicated in \cite{Bartels:1999aw}, the integral
equation (\ref{d6I}) does not represent the final version.
Instead, it will be necessary, once more, to separate a
reggeizing piece:
\bea
D_6^I = D_6^{I;R} + D_6^{I;I}
\label{2nd-reduction}
\eea
and rewrite the equation. Our ansatz for the first term,
$D_6^{I;R}$, will be constructed in
such a way that, to lowest order, it removes the $L$ and the
$I+J$ terms in (\ref{L-structure1}),
 (\ref{I-structure1}),and (\ref{J-structure1}). We put:
\bea
D_6^{I;R}=D_6^{I;R(I+J)}+ D_6^{I;R(L)}
\label{2nd-reduction-R}
\eea
with the two terms being illustrated in Fig.\ref{fig8}a and b:
\begin{figure}
\begin{center}
\epsfig{file=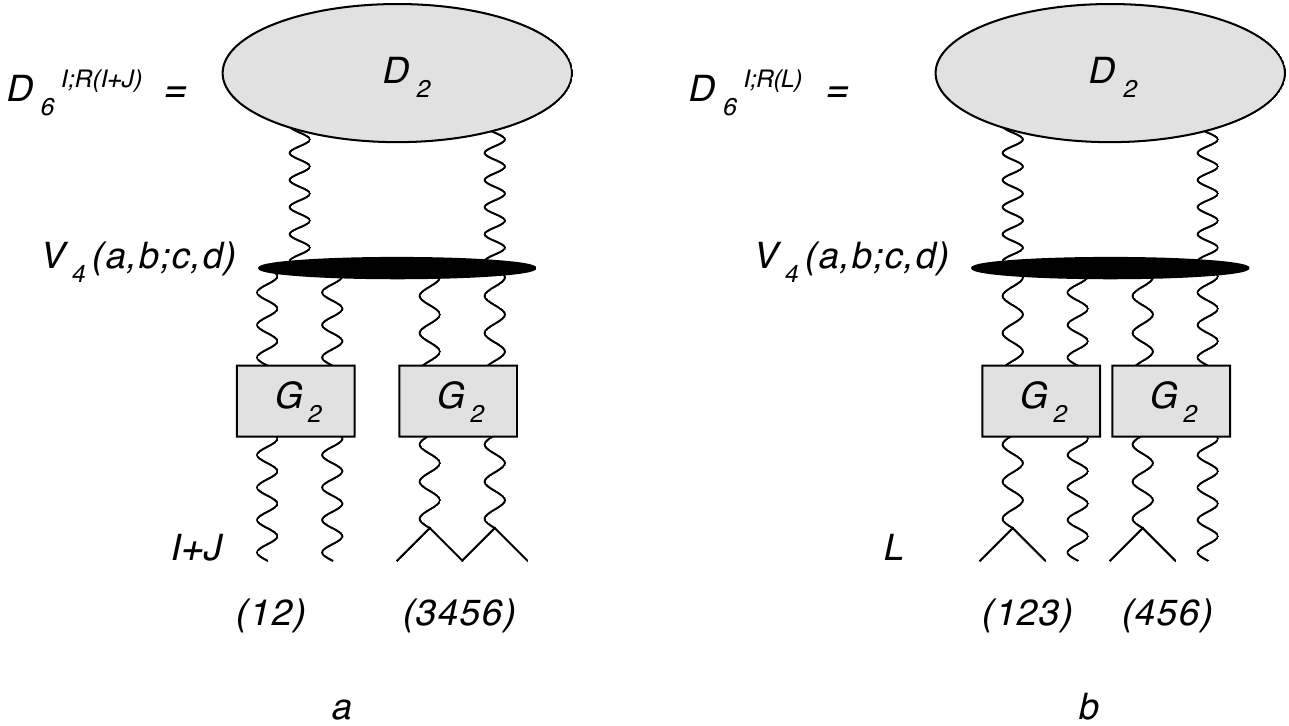,width=8cm,height=5cm}
\end{center}
\caption{Reggeizing pieces of the second reduction, $D_6^{I;R}$: (a) $D_6^{I;R(I+J)}$, (b) $D_6^{I;R(L)}$.
As in Fig.\ref{fig6}, the pairs of straight lines at the lower end of reggeized gluons indicate that they  depend upon the sum of momenta and finally decay into two gluons.   
Further details are described in section (4.4).}
\label{fig8}
\end{figure}
Fig.\ref{fig8}a belongs to the partition $(12)(3456)$, Fig.\ref{fig8}b to the
partition $(123)(456)$. Both terms contain the vertex $V_{4 \ot 2}$
which, according to (\ref{d4tot}), consists of three terms. But in the large-$N_c$ limit and for 
our color configuration, only the first term contributes, in
which the pairs (ab) and (cd) are color singlet states. Moving
further down in Fig.\ref{fig8}, we recognize two separate BFKL ladders;
this separation is a consequence of the large-$N_c$ limit. For
the first term, Fig.\ref{fig8}a, we have at the bottom the color and
momentum structure described in (\ref{I-structure1}) -
(\ref{J-structure2}). The 'L'-term in Fig.\ref{fig8}b, described in (\ref{L-structure1})- (\ref{L-structure3}), consists of products of two triplets, and in each triplet we have a sum of three terms.

We emphasize that the reggeizing pieces have been constructed in such a way that, to lowest order, they coincide with the second and third terms on the rhs of Fig.\ref{fig7}. Taking into account the subsequent large-$N_c$ evolution, this requirement fixes the reggeizing pieces uniquely. Further below (in section 4.4) we will come back to these reggeizing terms and discuss in more detail their significance.

\subsection{The large-$N_c$ limit of
Eq.(\ref{d6I})}

With the ansatz (\ref{2nd-reduction}) and (\ref{2nd-reduction-R}) we return
to (\ref{d6I}) (or Fig.\ref{fig7}) and insert it into the terms containing $D_6^{I}$:
\bea
&&S_6 \left( D_6^{I;R(L)} +D_6^{I;R(I+J)}+D_6^{I;I}\right)\nonumber\\
&&= W_{odd}D_2 +T_L D_2 +
(T_I+T_J)D_2 +
\sum K_{3 \ot 2} D_5^I+ \sum K_{4 \ot 2} D_4^I \nonumber\\
&& +\sum_{(ij)} V_{ij} \left( D_6^{I;R(L)}
+D_6^{I;R(I+J)}+D_6^{I;I}\right).
\label{d6II}
\eea
Our goal is the derivation of an equation for $D_6^{I;I}$.
As it was said before, in this paper we consider the
configuration where the six gluon state consists of three color
singlet pairs, and we consider the large-$N_c$ limit of the
corresponding six gluon amplitude. To be definite, we demand
that the pairs $(12)$, $(34)$, and $(56)$ are in color singlet
states. This allows for the three partitions $(12)(3456)$, $(1234)(56)$, and $(1256)(34)$.

\begin{figure}
\hspace*{0pt}
\epsfig{file=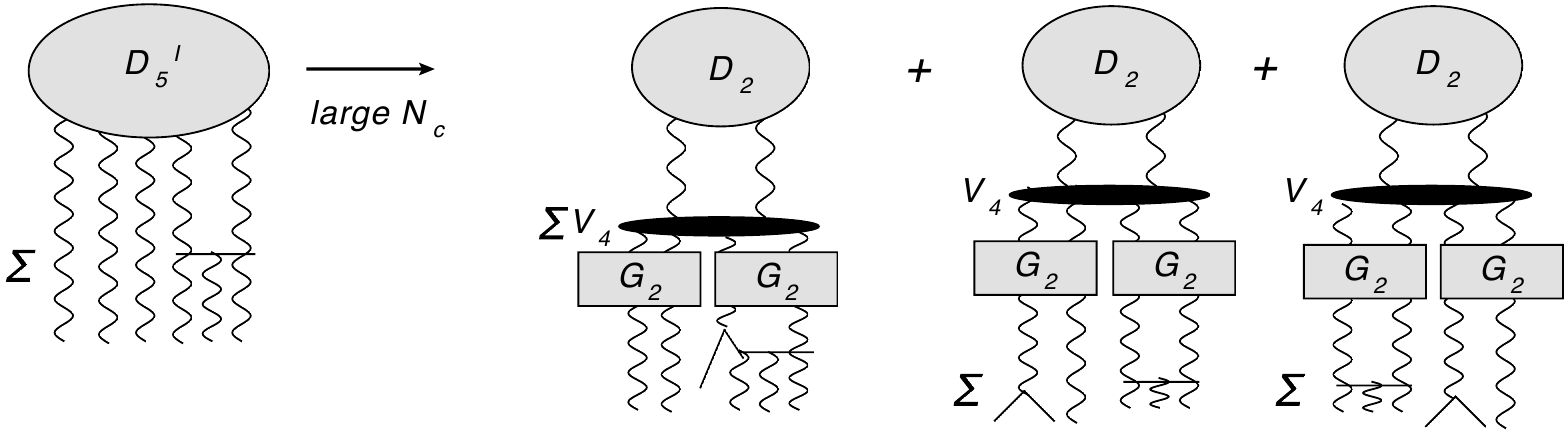,width=15cm,height=4cm}
\caption{The three terms emerging from the $D_5^I$ term in Fig.\ref{fig7}.
The first term belongs to the partition $(12)(3456)$, the second and third ones
to the partition $(123)(456)$.}
\label{fig9}
\end{figure}

We proceed by discussing, on the rhs of (\ref{d6II}), term by term. We will find that for our color configuration, in the  large $N_c$ limit the rhs simplifies. For reasons which will become clear soon we begin with
the pieces containing $D_5^Î$ and the $2 \to 3$ kernel. From (\ref{d5tot}) we see that $D_5^Î$ is a sum of 10 terms.
Each term contains a $V_{4 \ot 2}$ vertex. Below this vertex we have the evolution of the four reggeon state, which in the large-$N_c$ limit splits into the evolution of two independent two-reggeon states (\ref{G4-large-Nc}). At the end of this evolution one of the four reggeized gluons splits into two gluons. Subsequently, we attach a vertex with $K_{3 \ot 2}$. This structure implies that the $D_5$ term on the rhs of Fig.\ref{fig7} can be written as a sum of different groups of contributions: they are illustrated in Fig.\ref{fig9}.
In the first group we have the partition $(12)(3456)$, where the subsystem
$(12)$ evolves independently, and the $2\to3$ transition is
inside the subsystem $(3456)$. Second, in the partition
$(123)(456)$ there exists the part of $D_5^{I}$ which contains
the splittings $(12)$, $(23)$, and $(13)$, and the $2\to 3$
transition sits inside the triplet $(456)$. In the third group, the $2\to 3$
transition is inside the triplet $(123)$.
Consequently,  only the first group contributes to our
configuration of three color singlet pairs, whereas the last two terms have triplet structures.
As we will discuss below, they have to be combined with the $L$-terms and are not important for the following discussion.
Let us discuss the first group in more detail.  Beginning with $D_5^{I}$ in
 (\ref{d5tot}), a color singlet of the pair $(12)$ can come only from the last three terms:
\bea
\label{D5-first}
&&D_5^{I,a_1a_2a_3a_4a_t} \to \\
&&\frac{g}{2} \Big[f^{a_3a_4c} D_4^{I,a_1a_2ca_5}(1,2,34,5)+
f^{a_3a_5c} D_4^{I,a_1a_2ca_4}(1,2,35,4)+f^{a_4a_5c} D_4^{I,a_1a_2a_3c}(1,2,3,45)\Big]. \nonumber
\eea
In the first term we have
\beq
D_4^{I}(1,2,34,5)=(G_4 V_{4 \ot2} D_2)(1,2,34,5).
\eeq
As we have said before, the evolution in $G_4$ factorizes:
\beq
G_4(1,2,3,4)=G_2(1,2)G_2(3,4) ,
\eeq
and we arrive at:
\beq
G_2(1,2) G_2(34,5) (V_4 D_2) (1,2;34,5) \delta_{a_1a_2} f^{a_3a_4a_5},
\eeq
where the Green's functions $G_2(1,2)$ and $G_2(34,5)$ (suppressing the color factors) describe
the pomeron evolution in the subsystems $(1,2)$ and $(34,5)$ resp. Together with analogous expressions for the other two terms in (\ref{D5-first}) we find:
\bea
&&D_5^{I,a_1a_2a_3a_4a_t} \to \hspace{5cm}\nonumber\\
&&=\frac{g}{2} \delta_{a_1a_2} f^{a_3a_4a_5}
G_2(1,2) \Big[G_2(34,5)  (V_4 D_2) (1,2;34,5)-G_2(35,4)  (V_4 D_2) (1,2;35,4)\nonumber\\
&&+G_2(3,45)
(V_4 D_2) (1,2;3,,45) \Big].
\eea
Apart from the extra Green's function $G_2(1,2)$ we have the same structure as $D_3$  in (\ref{d31}). We attach the $2 \to 3$ vertex:
\bea
T_2&=& \sum \left(K_{3 \ot 2} D_5^{I}\right)\nonumber\\
&=& G_2(1,2)  \sum \left( K_{3 \ot 2} G_2(3',4')\right)  (V_4 D_2)(12,3',4').
\eea

As the next term on the rhs (\ref{d6II}) we consider the triplet states. The first term on the rhs of (\ref{d6II})
contains the pomeron$ \to 2$ odderon vertex. Its structure is explained in (\ref{Odd-vertex}): the two gluon state splits into two
triplet states which are characterized by the symmetric $d^{a_1a_2a_3}$ tensors. As an example, we once more chose the partition $(123)(456)$ at the $W$vertex in (\ref{Odd-vertex}). Below
this vertex, the two triplet states, in
the large-$N_c$ approximation, evolve separately and do not mix.
Each of them forms an QCD odderon state, as described for example, in
\cite{Bartels:1999yt,wj,jw}.

 A similar discussion applies to the
second term on the rhs of (\ref{d6II}), the $L$-term (cf. ((\ref{L-structure1}) - (\ref{L-structure3})): below the vertex $V_{4 \ot 2}$ two triplet states start,
this time accompanied with $f$-tensors.
As written in (\ref{L-structure2})-(\ref{L-structure3}), in each of the two triplets the momentum structure  is the same as for $D_{30}$ in (\ref{d30}).
These terms represent the inhomogeneous terms, the last two terms in Fig.\ref{fig9} the
kernels for the evolution equation of the reggeizing terms $D_6^{I;R(L)}$. They exactly equal the
$D_6^{I;R (L)}$ term on the lhs of the integral equation (\ref{d6II}):
\beq
S_6 D_6^{I;R(L)}=T_L D_2+  \sum K_{3 \to 2} D_5^Î |_{\text{last two terms in Fig.8}}
+\sum_{(ij)}  V_{ij} D_6^{I;R(L)}.
\eeq

We note that these triplet terms, at large $N_c$,  do not contribute to the color structure we are considering in this paper.
However we note that the terms with the f-structure will be crucial when we turn to the next step, the case of 8 gluons. Here we
expect new fan diagrams: in addition to the fan diagram in Fig.\ref{fig2} we should find a higher order fan diagram   where the lower BFKL
ladder on the lhs will split into two ladders. This is where the 'L' terms
will be needed for building the two lower triple pomeron vertices.

Returning to the rhs of (\ref{d6II}) we, from now on, disregard the odderon term on both sides of the equation: as an aditive contribution to $D_6^{I}$ it does not mix with the other pieces and satisfies its own equation. Similarly, we remove $D_6^{I;R(L)}$ on the both sides of the equation, together with the $T_L D_2$ terms and parts of the $D_5$ 
contributions on the rhs. We are thus left with the combination $ D_6^{I;R(I+J)}+D_6^{I;I}$, where $D_6^{I;I}$ now is without the two-odderon piece (for simplicity, we continue to use the same notation). 

Let us then take a closer look at the remaining pieces. We begin with the terms $T_I$ and $T_J$, (\ref{I-structure1}) - (\ref{J-structure2}) and focus on the partition $(12)(3456)$..
For this partition, the couplings at the vertex
$V_{4 \ot 2}$ have the color structure $\delta_{a_1a_2} d^{a_3a_4a_5a_6}$. and
the momentum structure described in (\ref{I-structure2}), (\ref{J-structure2}). Considering the four reggeons '3', '4', '5', and '6',
the structure of the 7 terms in the sum 'I' and 'J' is exactly the same as
for  $D_{40}$ in (\ref{d40}). Reggeons '1' and '2' play the role of spectators:
\beq
T_1=\left( (T_I+T_J)D_2\right)(1,2;3,4,5,6).
\eeq

Turning to the terms containing $D_4^(I)$ (the next-to-last term in Fig.\ref{fig7}), we note that
the evolution of the four gluon state in $D_4^{I}$ is the sum of three states,
$(12)(34)$, $(13)(24)$, and $(14)(23)$, and in each of them the
evolution is described by two separate BFKL Green's functions. Only the first
term belongs to the color singlet $(12)$, and consequently the $2 \to 4$
kernel has to be inside the quartet $(3456)$. In rapidity space this term can be written in the following form:
\beq
T_3=\int dy' G_2^{(12)}(y_1-y')K_{4 \ot 2}G_2^{(3'4')}(y_2-y') (V_4 D_2(y'))(1,2;3'4').
\eeq

\begin{figure}
\hspace*{0pt}
\epsfig{file=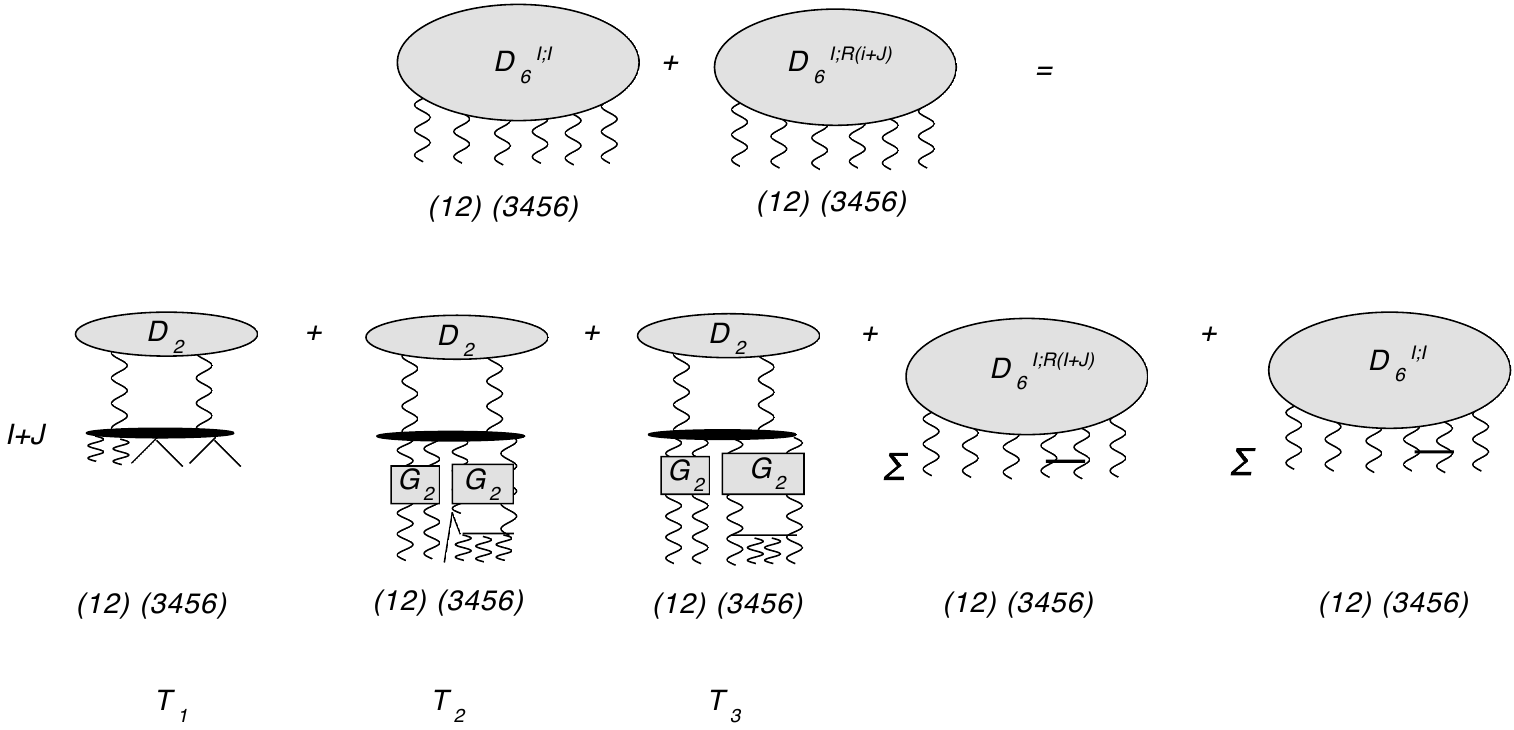,width=15cm,height=8cm}
\caption{The integral equation for the combination $D_6^{I;R(I+J)}+D_6^{I;I}$, restricted to
the partition $(12)(3456)$.
Here, in the last two terms, the sum over pairwise interaction includes
only the pair $(12)$ and pairwise interactions inside the quartet $(3456)$.}
\label{fig10}
\end{figure}

This finishes our discussion of the inhomogeneous terms on the rhs of  (\ref{d6II}), For the partition $(12)(3456)$ we therefore can write the following integral equation for the combination $D_6^{I;R(I+J)}+D_6^{I;I}$:
\beq
\label{d6II-sim}
S_6 \left( D_6^{I;R(I+J)}+D_6^{I;I}\right)= T_1+T_2+T_3+\sum_{(ij)} V_{ij} \left(D_6^{I;R(I+J)}+D_6^{I;I}\right).
\eeq
We illustrate this equation in Fig.\ref{fig10}.

\subsection{Two triple pomeron vertices}

Let us now analyze the integral equation illustrated in Fig.\ref{fig10}. Most strikingly, we notice
that the subsystem $(12)$ has been separated from the subsystem $(3456)$: the term
$D_6^{I;R(I+J)}$ on the lhs and on the rhs (illustrated in Fig.\ref{fig8}a), as well as the first three terms
in the second line of Fig.\ref{fig10} all contain $V_4$, and below we have a separate cylinder for the $(12)$ subsystem. To see this in more detail,
we first note that the reggeizing terms $D_6^{I;R(I+J)}$ (Fig.\ref{fig8}a)
in rapidity space can be written as
\beq
\int dy' G_2^{(12)}(y_1-y') \sum G_2^{(34,56)}(y_2-y') \left(V_4 D_2(y')\right)(1,2;3'4').
\label{D6-IR(I+J)}
\eeq
Here the sum in front of $G_2^{(34,56)}(y_2-y')$ stands for summing,at the lower end of the BFKL evolution, over momentum and color structures, in accordance with  (\ref{d40}).
Interpreting the rapidity evolution in (\ref{D6-IR(I+J)}, we compare  with Fig.\ref{fig8}a:
starting from above, we evolve up to rapidity $y'$ where the two-gluon state merges into the four gluon state. Below, the pair $(12)$
evolves up to $y_1$, whereas the other two-gluon state evolves up to $y_2$ where it splits into the four gluons $(3456)$.

Next, for the inhomogeneous terms $T_1+T_2+T_3$  we note that - formally - we could solve the integral equation (\ref{d6II-sim})
by introducing the Green's function $G_6$: in rapidity space we put
\beq
S_6=\partial_y+\sum_{i=1}^{6} \omega(i),
\eeq
and  (\ref{d6II-sim}) can be solved by
\beq
\left(S_6-\sum_{(ij)} V_{ij}\right) \left( D_6^{I;R(I+J)}+D_6^{I;I}\right) = T_1+T_2+T_3
\eeq
or
\beq
\label{d6I-sim}
\left( D_6^{I;R(I+J)}+D_6^{I;I}\right) = G_6 (T_1+T_2+T_3)
\eeq
with
\beq
\left(S_6-\sum_{(ij)} V_{ij}\right) G_6(y)=\delta(y).
\eeq
In the large-$N_c$ limit there is no interaction between the systems $(12)$ and $(3456)$, such that in
the Green's function we can simplify:
\beq
S_6-\sum_{(ij)} V_{ij}   \to S_6- V_{12} - \sum_{3 \le i,j} V_{(ij)}.
\eeq
Consequently, the Green's function $G_6$ splits into a product:
\beq
G_6(1,2,3,4,5) \to G_2(1,2) G_4(3,4,5,6),
\eeq
where each of the factors satisfies its own equation:
\beq
\left(S_2-V_{12} \right)  G_2(y;1,2)=\delta(y),\,\,\left( S_4-  \sum_{3 \le i,j} V_{(ij)} \right) G_4(y;3456) =\delta(y).
\eeq

We now use this decomposition to evaluate the three terms on the rhs of (\ref{d6I-sim}). For $T_1$ we find:
\bea
&&\int dy' G_2(y_1-y';12) G_4(y_2-y',3456) T_1(y')
\nonumber\\
 =&&\int dy' G_2(y_1-y';12) G_4(y_2-y',3456) (T_I+T_J)D_2(y').
\eea
Comparison with Fig.\ref{fig10} shows that the transition from 2 to 4 gluons occurs at rapidity $y'$.

For the other two terms, $T_2$ and $T_3$, the situation is a bit more complicated since
they contain already evolution inside the subsystem $(3456)$. Using the identity
\beq
\int dy' G_2(y_1-y') G_2(y'-y_2) = G_2(y_1-y_2)
\eeq
we write for the $T_2$ term:
\bea
&&\int dy' G_2(y_1-y';12) \left( \int dy'' G_4(y_2-y'';3456) \sum \left( K_{3 \ot 2} G_2(y''-y';(3+3')4')\right) \right) \cdot \nonumber\\ &&\hspace{6cm}\cdot(V_4 D_2(y'))(12,3'4').
\eea
Similarly, for the $T_3$ term
\bea
&&\int dy' G_2(y_1-y';12) \left( \int dy'' G_4(y_2-y'';3456) K_{4 \ot 2}G_2(y''-y';3'4') \right) \cdot \nonumber\\
&&\hspace{6cm}\cdot(V_4D_2(y'))(12,3'4').
\eea
Again, Fig.\ref{fig10} provides an illustration of these two equations.

All three terms have the common Green's function in the subsystem $(12)$. In order to fully separate the evolution of the pair $(12)$, we choose $y_1<y_2$, fix the rapidity  $y'=y_1$, and use the
property
\beq
G_2(0)=1,
\eeq
i.e. the pair $(12)$ does not evolve in rapidity.
This then leads to the following expression for the rhs of (\ref{d6I-sim}): 
\bea
\label{D6II-sim}
&&G_6(T_1+T_2+T_3)|_{y'=y_1}\\
&&= G_4(y_2-y_1;3456) (T_I+T_J)D_2(y_1)\nonumber\\
&&+\left( \int dy'' G_4(y_2-y'';3456) \sum \left( K_{3 \ot 2} G_2(y''-y_1;,(3+3')4')\right) \right)  (V_4 D_2(y_1))(12,3'4') \nonumber\\
&&+\left( \int dy'' G_4(y_2-y'';3456) K_{4 \ot 2}G_2(y''-y_1) \right) (V_4D_2(y_1))(12,3'4').\nonumber
\eea

This equation has to be compared with the solution of $D_4$ in eq.(\ref{sol-eqd4}). On the rhs of
(\ref{D6II-sim} we keep the momenta of the pair $(12)$ fixed and concentrate on the subsystem 
$(3456)$ below the upper triple pomeron vertex, i.e. $V_4D_2(y_1)$. Then we see, term by term, that the rhs of (\ref{D6II-sim}) is the same as of  (\ref{sol-eqd4}),
provided we substitute $D_{20} \to (T_I+T_J)D_2$.  The second and third lines on the rhs of (\ref{D6II-sim}) correspond to the second and third terms on the rhs of (\ref{sol-eqd4}), once we take into account that $\sum K_{3 \ot 2} D_3$ in (\ref{sol-eqd4}) really means the same as $\sum K_{3 \ot 2} G_2$ in (\ref{D6II-sim}).

We then make use of the discussion given after (\ref{sol-eqd4}). For the subsystem $(3456)$ below  $V_4D_2(y_1)$ we have the same structure as for $D_4$, i.e. we have the sum of reggeizing piece and an irreducible one.
The former one has the form:
\beq
 \sum G_2^{(34,56)}(y_2-y_1) \left(V_4 D_2(y_1)\right)(1,2;3'4'),
\eeq
where the symbol $\Sigma$ reminds that the Green's function $G_2^{(34,56)}(y_2-y_1)$ at its lower end has 
the same momentum structure as $D_4^R$.  The second term, the analogue of $D_4^{I}$, has the form:
\beq
\int dy'' G_2(y_2-y'';34) G_2(y_2-y'';56) V_4G_2(y''-y_1;3''4'')(V_4D_2(y_1))(12,3'4').
\eeq
Finally we re-introduce the integration of $y'$ and the Green's function for the pair $(12)$. The first part then takes the form 
\bea
&&\int dy' G_2^{(12)}(y_1-y') \sum G_2^{(34,56)}(y_2-y') \left(V_4 D_2(y')\right)(1,2;3'4')\nonumber\\
&&=D_6^{I;R(I+J)}(y_1,y_2)
\eea
and thus coincides with the reggeizing piece $D_6^{I;R(I+J)}$ in (\ref{D6-IR(I+J)}). The second term becomes:
\bea
\label{D6II}
&&D_6^{I;I}(y_1, y_2,y_3)=
\int dy' G_2(y_1-y';12)
\int dy'' G_2(y_2-y'';34) G_2(y_3-y'';56)\nonumber\\
&&\hspace{1cm}\cdot V_4G_2(y''-y';3''4'')(V_4D_2(y'))(12,3'4')
\eea
and has the fan structure illustrated in Fig.\ref{fig2}.
Therefore we conclude that the rhs of (\ref{D6II-sim}) which had been  obtained from evaluating the rhs of  (\ref{d6I-sim}) 
is, in fact, the sum
\beq
D_6^{I}=D_6^{I;R(I+J)}+D_6^{I;I} 
\eeq
with $D_6^{I;R(I+J)}$ in (\ref{D6-IR(I+J)}) and $D_6^{I;I}$ in (\ref{D6II}).
The fan structure in eq.(\ref{D6II}) (Fig.\ref{fig2}) represents the main result of this paper.

\begin{figure}
\hspace*{0pt}
\begin{center}
\epsfig{file=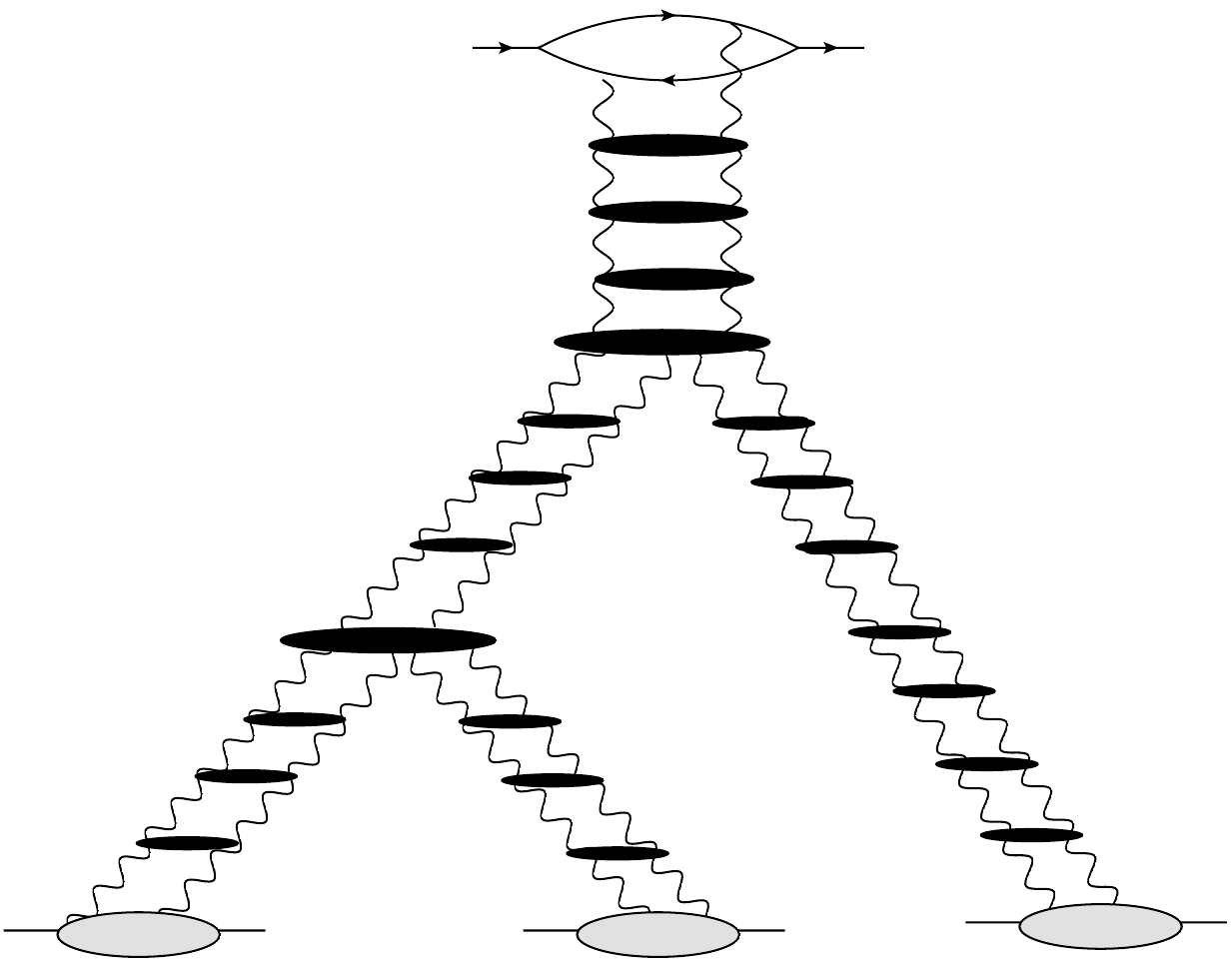,width=6cm,height=5cm},\epsfig{file=fan.pdf,width=6cm,height=5cm}
\end{center}
\caption{Two more fan diagrams}
\label{fig11}
\end{figure}

For completeness a few words need still to be said about the
remaining two partitions,
$(1234)(56)$ and $(1256)(34)$. By going through the same
sequence of arguments
as described for the partition $(12)(3456)$ one arrives at the
fan diagrams shown in
Fig.\ref{fig11}. It is only after adding these two additional fan
diagram that the Bose symmetry of
pomeron diagrams is restored.

\subsection{Reggeizing pieces in the four and six giuon amplitudes $D_4$ and $D_6$}

Finally we want to comment on the reggeizing terms, $D_4^R$, $D_6^R$, and $D_6^{I;R(I+J)}$ and $D_6^{I;R(L)}$. In our derivation, they appear as additive corrections to the 'simple' fan diagrams, and they require an interpretation. 

Let us first return to $D_4^R$, which we have discussed already at the end of section 2.2.  First a word about the order of this term. Compared to $D_2$, $D_4^R$   is of a higher order in the coupling constant : it contains an extra $\alpha_s$. Projecting  onto the target color
configuration where the two pairs $(12)$ and $(34)$ form singlets,
it has the order
$\alpha_s^2N_c^3(1+\sum_{n=1}\alpha_s N_cy)$ where the sum is provided by evolution.
This can be compared to the contribution from the triple pomeron, which has the order
$\alpha^3N_c^4y(1+\sum_{n=1}\alpha_s N_cy)$ where again the sum is given by evolution.
As one observes these two contributions are of the same order provided $\alpha_sN_cy\sim 1$
which is assumed in the leading logarithmic approximation.
Separating the contribution from the impact factor of the order $\alpha_sN_c$ one observes that the
triple pomeron without evolution has the same order as the reggeized term with a single interaction
inside the BFKL chain.

Next we recapitulate our interpretation of $D_4^R$. Starting from its simplest form, the coupling of a BFKL pomeron to the nucleus is given by the coupling to a single nucleon inside the nucleus, and this coupling is described by the gluon density of the nucleon. Following this picture,
$D_4^R$ in Fig.\ref{fig6}b describes the coupling of the pomeron to a pair of nucleons, and in   ~\cite{braun:tmf} it has been shown that this can be intepreted as the second term in an eikonal expansion of the initial conditions of the nonlinear evolution equations. In the original derivation of the BK equation ~\cite{Kovchegov:1999yj}
the low-energy scattering of the $q\bar{q}$ loop on the nucleus in the nucleus rest frame
was taken in the natural eikonal form, following the lowest order (classical) picture
\cite{Kovchegov:1998bi}. The interpretation of $D_4^R$ given in
\cite{braun:tmf} clearly corresponds to this picture.

However the validity of this picture also for the projectile proton is not so obvious,
especially when one wants to apply the saturation momentum approach to the proton itself.
So in practical calculations for hA and AB scattering within the JIMWLK approach
certain  more phenomenological initial conditions were taken, guided by the experimental data, e.g. in
 ~\cite{dusling, dumitru1, dumitru2}. They do not eikonalize the scattering on
the nucleus in terms of the scattering on the proton at low energies, and so they imply a non-standard spatial picture of the large heavy nucleus as composed  of individual nucleons.

For completeness we have to include into our discussion also $D_3$ (Fig.\ref{fig4}) and $D_5^R$ . In $D_3$, one of the lower reggeons decays into two elementary gluons: we interpret this as a higher order correction to the lowest order coupling of the BFKL pomeron to a nucleon. Similarly, $D_5^R$ has the same structure as $D_4^{I}$, where one of the two lower pomerons has the properties of $D_3$. 

\begin{figure}
\hspace*{0pt}
\begin{center}
\epsfig{file=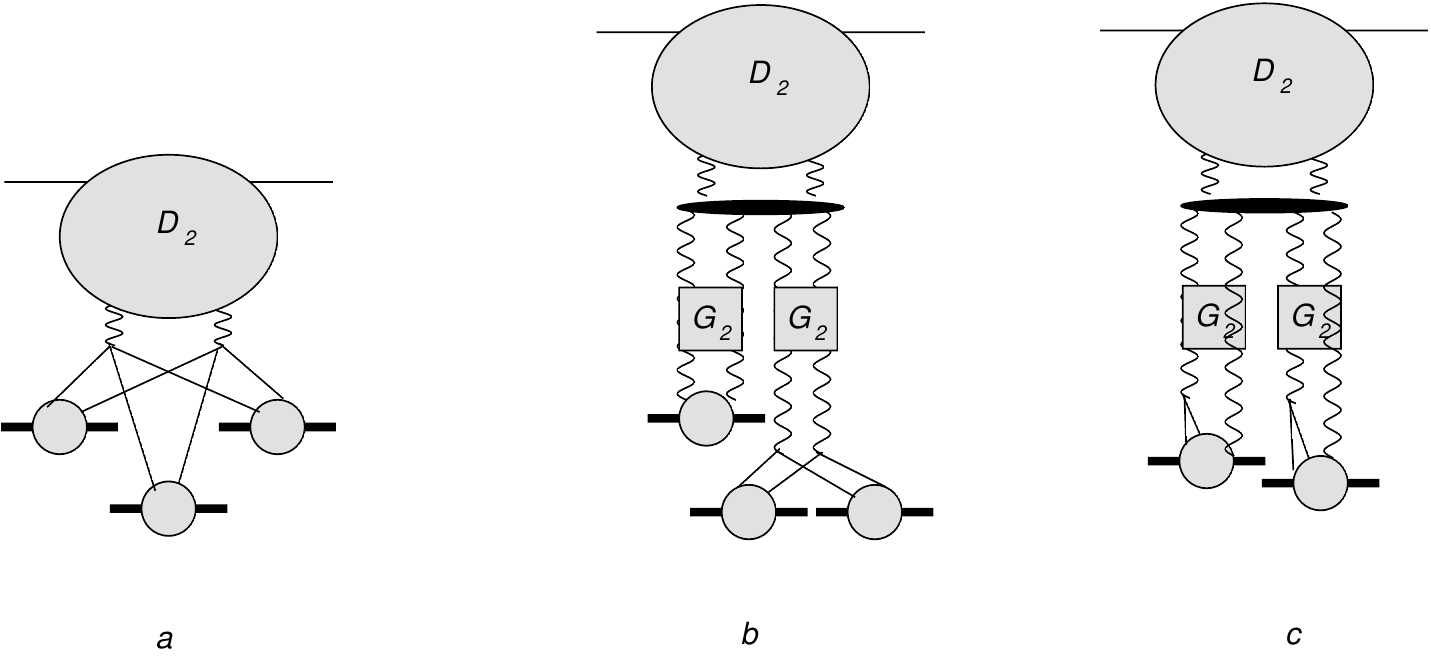,width=12cm,height=5cm}
\end{center}
\caption{Reggeized terms in $D_6$: (a) a term in $D_6^R$ in (\ref{decomp6}); (b) a term in  $D_6^{I;R(I+J)}$;(c) a term in $D_6^{I;R(L)}$.}
\label{fig12}
\end{figure}

In $D_6$ we again have found reggeizing terms:  $D_6^R$ in (\ref {decomp6}), and $D_6^{I;R}$ in (\ref{2nd-reduction}) which, in fact, consisted of the two classes of contributions (\ref{2nd-reduction-R}), $D_6^{I;R(I+J)}$ and $D_6^{I;R(L)}$. The first term,  $D_6^R$ with the momentum structure given in (\ref{d60}), belongs to a single BFKL pomeron which decays into six elementary gluons and thus contains  couplings to (up to) three nucleons. 
As an example we illustrate in Fig.\ref{fig12}a the term, $D_2(135,246)$.
This term has to been seen in the context of $D_4^R$: the pomeron now couples to a triplet of nucleons: this supports our previous finding that reggeizing terms provide a nontrivial structure 
for the couplings of pomerons to the nucleus.   

The detailed structure of $D_6^{I;R(I+J)}$ is contained in (\ref{I-structure2}) and (\ref{J-structure2}), the structure of  $D_6^{I;R(L)}$ in (\ref{L-structure3}). Again, we illustrate examples. For the $I+J$ structure we choose, in Fig.\ref{fig12}b, the term with the momentum structure $(1,2;35,46)$. As one can see from  (\ref{I-structure2}) and (\ref{J-structure2}), the right BFKL pomeron at the lower end has exactly the same structure as the single pomeron in $D_4^R$ and thus couples to the nucleus in the same 
manner as the single pomeron in $D_4^R$. In Fig.\ref{fig12}b we show the analogue of Fig.\ref{fig6}b, the coupling to a pair of 
nucleons. As to the L-structure, a look at (\ref{L-structure3}) shows that now each of the pomerons at 
the lower end has the structure of $D_3$: an example is shown in Fig.\ref{fig12}c.  
In summary, both sets of reggeizing contributions, $D_6^{I;R(I+J)}$ and $D_6^{I;R(L)}$, have the same structure as $D_4^{I}$, but the two lower pomerons are corrected: either both pomerons decay into three gluons and have the structure of $D_3$, or one of them decays into four gluons and repeats $D_4^R$ whereas the other one remains uncorrected.

\begin{figure}
\hspace*{0pt}
\begin{center}
\epsfig{file=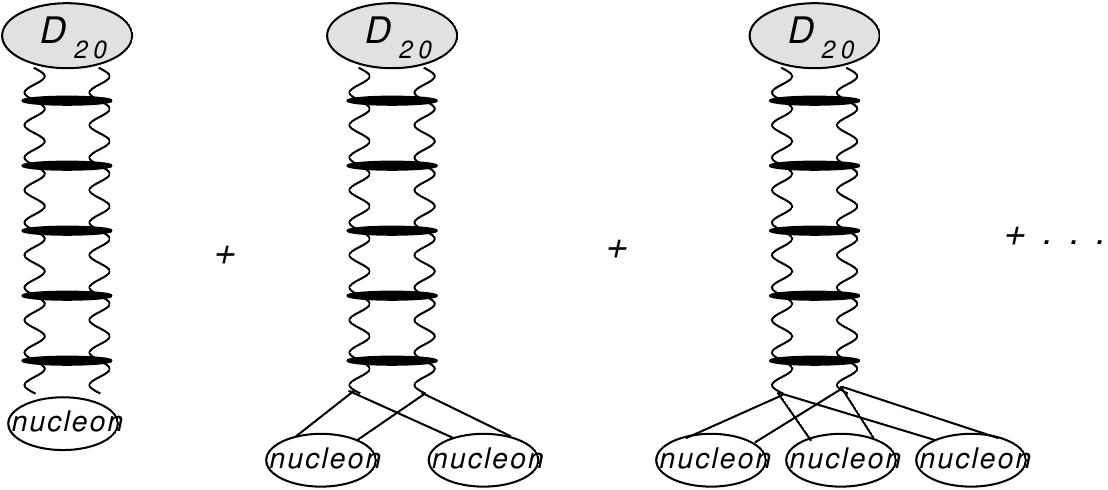,width=10cm,height=4cm}\\ 
\vspace{0.5cm}
\epsfig{file=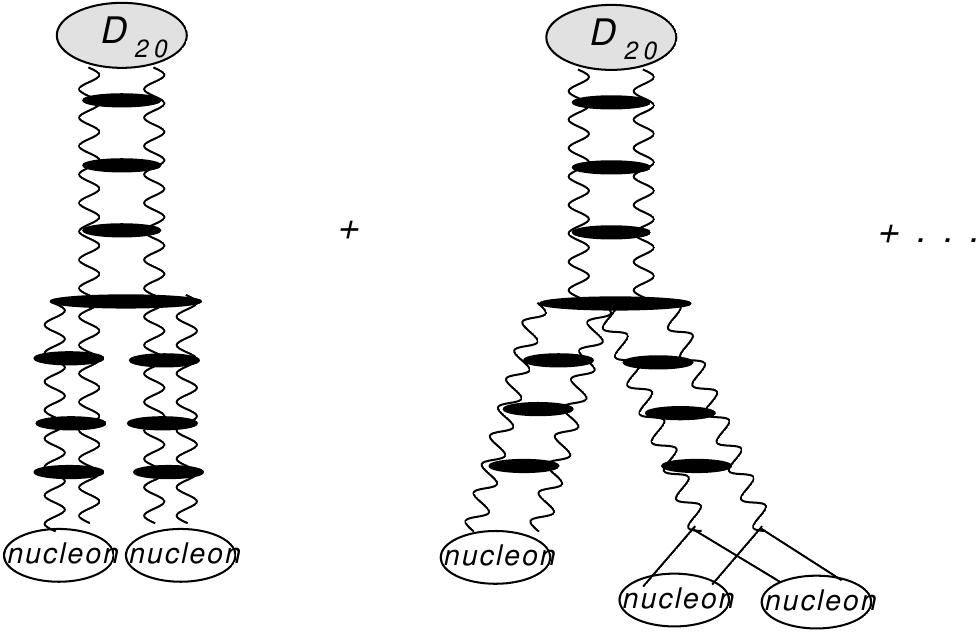,width=8cm,height=4cm}\\
\vspace{0.5cm}
\epsfig{file=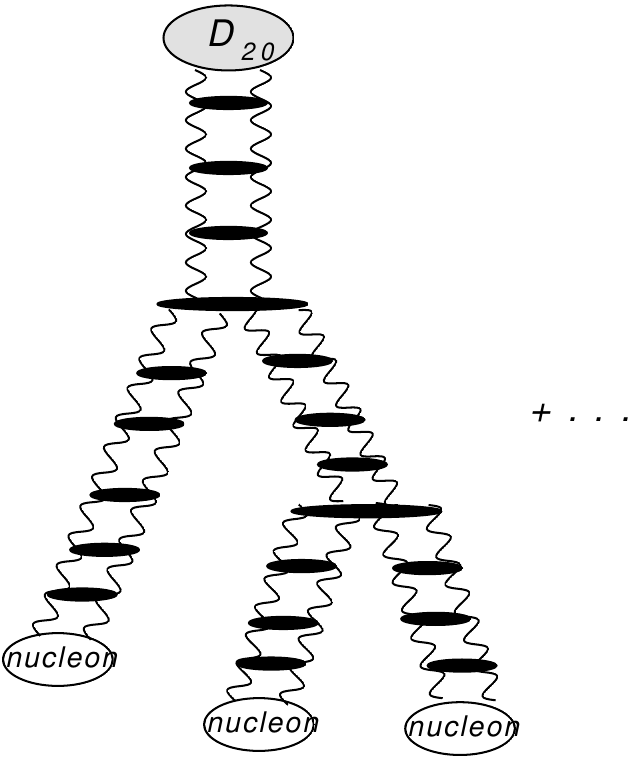,width=6cm,height=4cm}\\
\end{center}
\caption{Illustrating our main results: in the upper line the single pomeron can couple to one, two, three nulceons;
in the second line we show the first fan diagram with couplings to one, two nucleons;
in the third line we have the second order fan diagrams with couplings to single nucleons}
\label{fig13}
\end{figure}

Let us summarize our main findings for $D_2$, $D_4$ and $D_6$ in Fig.\ref{fig13} (see also Fig.\ref{fig6}b and \ref{fig12}a). In addition to the 'simple' fan structure (on the lhs) where each pomeron at the lower end couples to a single nucleon inside the nucleus, we found reggeizing pieces (on the rhs) which
are interpreted as coupling to two or more nucleons and can be absorbed into suitable intitial condtions    
	    
These results suggest the following generalization to amplitudes $D_{2n}$ with $2n$ reggeons at the lower end: in addition to the fan diagrams with $n$ pomerons at the lower end we should find a hierarchy of reggeizing pieces: single pomerons which decay into $2n$ elementary gluons and thus can couple to up to $n$ nucleons, the first fan diagram with two pomerons which can decay into $(2,2(n-2))$, $(3,(2n-3))$ gluons etc.      

We consider the existence of our reggeizing pieces, together with their interpretation, as quite an important result of our analysis. Namely, such contributions related to the reggeization of the gluons do not show up in derivations based upon the dipole picture. However, a closer look at the derivation of the BK-equation \cite{Kovchegov:1999yj} shows that the situation is more subtle. Namely, in addition to the nonlinear evolution which is derived within in the dipole picture, a specific form of the coupling to nuclei is also included; in coordinate space this coupling has an  eikonal structure, and when expanded it describe the coupling to one, two, three nucleons, similar to the first line of Fig.\ref{fig13}. From what we can say at the moment (but as we have emphasized before, the detailed analysis of our couplings still needs some further investigation), both approaches seem to arrrive at the consistent  result:\\
- our derivation (performed in perturbation theory in momentum space) leads to contributions which arise from the reggization of the gluon but can be reformulated as specific corrections in the couplings of the 
pomerons to the nuclei;\\
- the derivation based upon the color dipole picture does not see such 'reggeized' contributions, but by including the nontrivial form of couplings of pomerons to the nuclei it leads to the same result. 

Clearly, a more detailed analysis of this equivalence is highly desirable and we hope to come back to this in a future publication.


\section{Discussion}

In our study we have used the framework of QCD field theory of
reggeized gluons for discussing, in the large-$N_c$ limit, diagrams with up to six
reggeized gluons, using the results of \cite{Bartels:1999aw}.
Our main interest was the formation of two
consecutive triple pomeron vertices. We
have found that in diagrams with higher splitting kernels
$K_{n \ot 2}$ with $n=$5 and 6 all those contributions are
present which are needed to form the triple pomeron vertex at
each splitting. This result agrees with the conclusions derived
in the framework of the dipole model, the Balitski-Kovchegov
equation.

In the course of our analysis a few other terms appeared, which require some attention.
First, our analysis contains
the pomeron $\to$ two odderon vertex (see the discussion in section 3, after (\ref{Odd-vertex})). 
This vertex was found as early as in 1999, when only the
Janik-Wosiek solution (JW) for the odderon \cite{wj,jw} was known, describing a C-odd bound state of three reggeized gluons.  Soon afterwards a new solution with intercept one was found
in \cite{Bartels:1999yt} (the BLV odderon). This solution has a simple quasi-two-gluon structure corresponding to the fusion of two of the three reggeized gluons into a single object (the zero range "digluon").
Since the structure of this odderon is practically identical to the pomeron, it became possible to
radically simplify vertices for both pomeron $\to$two odderon and odderon $\to$ odderon+pomeron transitions,
which actually reduce to the triple pomeron form. This allowed to easily derive the coupled
system of evolution of the  pomeron and odderon including both transitions ~\cite{kovchnew,itakura}.
This system was analyzed in rather crude approximation in ~\cite{Motyka:2005ep} where the angular dependence was drastically simplified. Recently this very system was
investigated in detail in the JIMWLK approach and solved numerically, albeit without the pomeron $\to$
two odderon transition
~\cite{lappi}. We find it important to stress that the pomeron$\to$ two odderon vertex found in \cite{Bartels:1999aw} and in our present analysis, describing the full transition from 2 to 6 reggeized gluons, is not restricted to the BLV form of the odderon but is valid also for a more general odderon consisting of three gluons, such as the JW odderon.
It would be interesting to study the coupled evolution of pomerons and odderon with the full angular dependence and the JW odderon as well.

Second, our investigation has led us  to separate, from the pomeron fan diagrams,
an extra class of contributions which result from the reggeization of the gluon ('reggeized terms'). 
A detailed discussion has been given in section 4.4 and the main results are summarized in Fig.\ref{fig13}. 
The first example, for the case of four reggeized gluons, was $D_4^R$ (Fig.\ref{fig6}b). For the six-gluon case
we first found $D_6^R$ (see (\ref{decomp6}). As illustrated in the first line of Fig.\ref{fig13} (see also Fig.\ref{fig6}b and Fig.\ref{fig12}a), both $D_4^R$ and $D_6^R$, contain the two gluon state of the pomeron, $D_2$, which at the lower end splits into
four or six gluons, resp. To be complete we have to include $D_3$ (defined in (\ref{d31})) and $D_5^R$ (given in (\ref{d5R})),  where $D_2$ splits into three or 5 gluons, resp.
Within a nucleus, all these contributions can be interpreted as couplings of a single pomeron to pairs, triplets,... of nucleons inside the nucleus. Somewhat later, Fig.\ref{fig8}, we encountered  new reggeizing pieces, $D_6^{I;R}$. 
As an example, $D_6^{I;R(I+J)}$ in Fig.\ref{fig8}a has the form of the lowest order fan diagram (second line of Fig.\ref{fig13}), with one of the two lower pomerons splitting into 4 gluons. 
It thus seems as if the reggeizing pieces, being fully compatible with the fan structure, start to 'dress' the lower order fan diagrams, by introducing nontrivial structures into the coupling of pomerons to the nucleus and thus 
leading to a result compatible with  \cite{Kovchegov:1999yj} .  

Finally, our investigation of the appearance of more than one triple
pomeron vertex within the framework of
QCD reggeon field theory based upon reggeized gluons also might
serve other purposes.
First, it would be interesting to analyze the six gluon system
also for finite $N_c$. Most likely, a $2\to 4$ reggeon vertex
for non-singlet color quantum numbers should exist;
there is also the possibility of a $4\to 6$ reggeon vertex. Our
large-$N_c$ study might provide help in such an analysis. More
general, experience shows that reggeon field theory interaction
vertices - either interactions between reggeized gluons or
interactions between bound state fields such as
the pomeron or the odderon - cannot directly be read off from
high energy Feynman diagrams. In the derivation based upon
energy discontinuities several steps of 'reduction' have to be
performed, as we have demonstrated in this paper. A very
attractive way
is the use of Lipatov's effective action where the reggeized
gluon appears as a fundamental degree of freedom. But also in this
framework interactions between reggeized gluons arise only after
a careful analysis of several contributions. Examples are
the transition vertex $1 \to 3$ reggeized gluons which has been
derived in \cite{Hentschinski:2009zz}. It has also been shown
that, in accordance with signature conservation, the vertex $1
\to 2$ reggeized gluons disappears.

An exciting open problem is the appearance of two triple pomeron
vertices in the scattering
of two nuclei on two nuclei. We hope to address this problem in
a future investigation.
\vspace{1cm}

\noindent
{\bf Acknowledgements}: One of us (MB) thanks Hamburg University
and the II.Institute
of Theoretical Physics for the hospitality. One of us (JB)
gratefully acknowledges stimulating
and helpful discussions with C.Ewerz.



\begin{thebibliography}{100}

\bibitem{Bartels:1999aw}
  J.~Bartels and C.~Ewerz,
  JHEP {\bf 9909} (1999) 026
  [hep-ph/9908454].

\bibitem{Bartels:1993ih}
  J.~Bartels,
  Z.\ Phys.\ C {\bf 60} (1993) 471.
  

\bibitem{Bartels:1994jj}
  J.~Bartels and M.~Wusthoff,
  Z.\ Phys.\ C {\bf 66} (1995) 157.
  


\bibitem{mueller}A.H.Mueller, Nucl. Phys. {\bf B 415} (1994)
373;
{\bf B 437} (1995) 107; A.H.Mueller and B.Patel, Nucl. Phys.
{\bf B 425} (1994) 471
%

\bibitem{Braun:1997nu}
  M.~A.~Braun and G.~P.~Vacca,
  Eur.\ Phys.\ J.\ C {\bf 6} (1999) 147
   [hep-ph/9711486].

\bibitem{Bartels:2004ef}
  J.~Bartels, L.~N.~Lipatov and G.~P.~Vacca,
  Nucl.\ Phys.\ B {\bf 706} (2005) 391
    [hep-ph/0404110].

\bibitem{Balitsky:1995ub}
  I.~Balitsky,
  Nucl.\ Phys.\ B {\bf 463} (1996) 99
    [hep-ph/9509348].


\bibitem{Kovchegov:1999yj}
  Y.~V.~Kovchegov,
  Phys.\ Rev.\ D {\bf 60} (1999) 034008
   [hep-ph/9901281].

\bibitem{bar2}
  J.~Bartels, M.~G.~Ryskin and G.~P.~Vacca,
  Eur.\ Phys.\ J.\ C {\bf 27} (2003) 101
   [hep-ph/0207173].

%
\bibitem{pech}
  R.~B.~Peschanski,
  Phys.\ Lett.\ B {\bf 409} (1997) 491
   [hep-ph/9704342].

\bibitem{bra}
  M.~A.~Braun,
  Eur.\ Phys.\ J.\ C {\bf 63} (2009) 287
    [arXiv:0901.3660 [hep-ph]].


%
\bibitem{bratar}
  M.~A.~Braun and A.~N.~Tarasov,
  Phys.\ Lett.\ B {\bf 726} (2013) 300
    [arXiv:1304.3014 [hep-ph]]




\bibitem{Braun:2000wr}
  M.~Braun,
  Eur.\ Phys.\ J.\ C {\bf 16} (2000) 337
    [hep-ph/0001268].
%
\bibitem{kovchnew} Y.V.Kovchegov, L.Szymanowski, S.Wallon, Phys. Lett. {\bf B 586} (2004) 267.
%
\bibitem{itakura} Y.Hatta, E.Iancu, K.Itakura, L.McLerran, Nucl. Phys. {\bf A 760} (2005) 172.
%






\bibitem{Braun:1997gm}
  M.~Braun,
  Eur.\ Phys.\ J.\ C {\bf 6} (1999) 321
   [hep-ph/9706373].


\bibitem{braun:tmf} M.A.Braun, Theor. and Math. Phys. {\bf 148} (2006) 923.

%


\bibitem{Bartels:1999yt}
  J.~Bartels, L.~N.~Lipatov and G.~P.~Vacca,
  Phys.\ Lett.\ B {\bf 477} (2000) 178
   [hep-ph/9912423].

\bibitem{wj}  J. Wosiek, R. A. Janik, Phys. Rev. Lett. 79, (1997) 2935.
%
\bibitem{jw} R. A. Janik,  J. Wosiek, Phys. Rev. Lett. 82, (1999) 1092

\bibitem{Kovchegov:1998bi}
  Y.~V.~Kovchegov and A.~H.~Mueller,
  Nucl.\ Phys.\ B {\bf 529} (1998) 451
    [hep-ph/9802440].

%
\bibitem{dusling}  K.Dusling, F. Gelis, T.Lappi, R. Venugopalan, Nucl. Phys., {\bf A 836} (2010) 159.
%
\bibitem{dumitru1} A.Dumitru, J.Lalilian-Marian, E.Petresku. Phys. Rev. {\bf D 84} (2011) 014018.
%
\bibitem{dumitru2} A.Dumitru, E.Petresku, Nucl. Phys. {\bf A 879} (2012) 59.







%
%







\bibitem{Motyka:2005ep}
  L.~Motyka,
  Phys.\ Lett.\ B {\bf 637} (2006) 185
   [hep-ph/0509270].
%
\bibitem{lappi} T.Lappi, A.Ramnath, K.Rummukainen, H.Weigert, Phys. Rev. {\bf D 94} (2016) 054014.


\bibitem{Hentschinski:2009zz}
  M.~Hentschinski,
  [hep-ph/0908.2576] .
\end{thebibliography}
\end{document}